\documentclass[iop, apj]{emulateapj}
 
\usepackage{epstopdf}
\usepackage{epsfig}
\usepackage[section] {placeins}

\begin{document}


\defcitealias{Thilker07}{T07}


 \title{Resolved HI Imaging of a Population of Massive HI-Rich Galaxies with Suppressed Star Formation}
 
 \author{
 Jenna J. Lemonias\altaffilmark{1},
David Schiminovich\altaffilmark{1},
Barbara Catinella\altaffilmark{2},
Timothy M. Heckman\altaffilmark{3},
Sean M. Moran\altaffilmark{4}
}
 
 \altaffiltext{1}{Department of Astronomy, Columbia University, 550 West 120th Street, New York, NY 10027, USA; jenna@astro.columbia.edu}
 \altaffiltext{2}{Swinburne University of Technology, Centre for Astrophysics and Supercomputing, P.O. Box 218 - Mail H30, Hawthorn, VIC 3122, Australia}
 \altaffiltext{3}{Department of Physics and Astronomy, The Johns Hopkins University, 3400 N. Charles Street, Baltimore, MD 21218, USA}
  \altaffiltext{4}{Smithsonian Astrophysical Observatory, 60 Garden Street, Cambridge, MA 02138, USA}
 

%
%
%

 
%

%
 
  
\begin{abstract}

Despite the existence of well-defined relationships between cold gas and star formation, there is evidence that some galaxies contain large amounts of HI that do not form stars efficiently.  By systematically assessing the link between HI and star formation within a sample of galaxies with extremely high HI masses (log M$_{HI}$/M$_{\odot}$ $>$ 10), we uncover a population of galaxies with an unexpected combination of high HI masses and low specific star formation rates that exists primarily at stellar masses greater than log M$_{*}$/M$_{\odot}$ $\sim$ 10.5.  We obtained HI maps of 20 galaxies in this population to understand the distribution of the HI and the physical conditions in the galaxies that could be suppressing star formation in the presence of large quantities of HI.  We find that all of the galaxies we observed have low HI surface densities in the range in which inefficient star formation is common.  The low HI surface densities are likely the main cause of the low sSFRs, but there is also some evidence that AGN or bulges contribute to the suppression of star formation.  The sample's agreement with the global star formation law  
highlights its usefulness as a tool for understanding galaxies that do not always follow expected relationships.


\end{abstract}

 
 


\section{INTRODUCTION}


Many theories of galaxy evolution assume that a galaxy's evolutionary stage and its cold gas content are tightly linked.  For example, it is often assumed that a depleted cold gas reservoir can explain why massive galaxies are redder and exhibit less active star formation, but recent observations have challenged this picture by identifying populations of galaxies that contain significant amounts of cold gas despite their low or moderate star formation rates \citep{Morganti2006, Serra2012, Young2013}.  Some attempts have been made to systematically search for these galaxies to understand them as a population.  In this paper we pursue this line of reasoning by studying a sample of massive galaxies with an unusual combination of high HI masses and low specific star formation rates.  



Galaxies that have more cold gas than expected can be placed into two broad categories: red, early-type galaxies with unexpected measurable HI and galaxies that have significantly more HI than their star formation rates (SFRs) would suggest.  In the first category, early-type galaxies were expected to be devoid of cold gas because they lack the spiral arms that indicate recent star formation.  The first systematic imaging search for HI in early-type galaxies, motivated in part by the evidence for HI in individual early-types, found that HI is common at low levels in early-type galaxies in the field and could play an important role in their evolution \citep{Morganti2006}.  More recent surveys have revealed that at least 32\% of such galaxies have detectable HI and that 20\% of ETGs have regularly rotating disks of HI \citep{Serra2012}.  \citet{Young2013} show that even red early-type galaxies have cold gas at detection rates of 10-34\% and that the detection rates are highest among the most massive (log M$_*$/M$_{\odot}$ $>$ 10.5) red early-types.  
Galaxies in the \citet{Serra2012} sample contain up to 10$^{10}$ solar masses of HI.  Because galaxies with more than this amount of HI are rare, study of these galaxies has so far focused on individual galaxies rather than entire populations.  

Galaxies with low rates of star formation compared to their cold gas content comprise a diverse population of not just red early-type galaxies but also massive spirals that are expected to contain cold gas but have a high gas content incommensurate with relatively low SFRs.  The reasons for their low SFRs can be difficult to ascertain.  Recent simulations and observations have attempted to determine which physical processes or conditions can prevent cold gas from collapsing and forming stars.  In simulations the heating and disruption of gas by AGN feedback are frequently used to explain the low specific star formation rates (sSFR = SFR/M$_{*}$) of massive galaxies \citep[e.g.][]{Croton2006, Hopkins2008, Gabor2011}.  \citet{Gabor2011} showed that feedback from radio-mode AGN can shut down star formation, but it is unclear how this would affect the cold gas reservoir \citep{Ho2008, Fabello2011b}.  \citet{Martig2009} show that morphological quenching can quench star formation in bulge-dominated galaxies without removing or heating the cold gas. 

A high cold gas content may coexist with a low SFR in galaxies such as those with inefficient star formation in their outer disks.  The outer disks could harbor large quantities of gas but observations have shown that their densities are low and typically dominated by HI over H$_2$ \citep[e.g.][]{Wyder2009, Bigiel2010}.  Inefficient star formation is frequently attributed to a low gas surface density below a threshold level of 3-10 M$_{\odot}$ pc$^{-2}$ that is required for H$_2$ to form and for star formation to proceed efficiently \citep{Schaye2004, Bigiel2008}.  Below this threshold level the SFR surface density decreases steeply; this downturn could be related to H$\alpha$-derived star formation thresholds in outer disks such as those discovered by \citet{Martin2001}.


Extended UV (XUV) disks and giant low surface brightness galaxies (GLSBs) are both populations of galaxies characterized by inefficient star formation.  XUV-disks are defined by the presence of low-level UV flux beyond the main stellar disk.  The prevalence of XUV-disks suggests that inefficient star formation in the outer disks of galaxies may be common \citep{Thilker2007}, especially around massive, bulge-dominated galaxies \citep{Lemonias2011}.  GLSBs have a less stringent definition than XUV-disks, but are generally known to have low surface brightness disks at optical wavelengths and massive HI disks that extend well beyond the optical radius.  Malin 1 \citep{Pickering1997} is the prototypical GLSB, and several similar galaxies have been discovered \citep[e.g.][]{Morganti1997, Portas2010}.  Simulations have shown that extended low surface density gaseous disks may be common if the cold gas building up the gaseous disk has a high angular momentum \citep{Kimm2011, Stewart2011, Lu2014}.

It has become clear that cold gas and star formation are not always linked in obvious ways.  To gain a deeper understanding of the role HI plays in galaxy evolution, both within the general population and in populations that deviate from the general population, it is necessary to have large, well-defined samples of galaxies with known star formation and gas properties.  Recent large HI surveys such as the GALEX Arecibo SDSS Survey \citep[GASS;][]{Catinella2010} and the Arecibo Legacy Fast ALFA Survey \citep[ALFALFA;][]{Giovanelli2005} complement GALEX (Galaxy Evolution Explorer) and SDSS (Sloan Digital Sky Survey) data on star formation and galactic structure by providing key measurements of the cold gas in galaxies.  The advent of large HI surveys such as these provides the first opportunity for us to select and study volume-limited populations of galaxies defined only by their gas content.


In this paper we take advantage of these recent HI surveys to assess the link between cold gas and star formation in a unique and systematic way.  Although the term cold gas generally refers to both atomic and molecular gas, we only have information about HI so we refer to cold gas and HI interchangeably throughout the rest of the paper except when noted otherwise.  We define a sample of massive (log M$_{*}$/M$_{\odot}$ $>$ 10.0) HI-rich galaxies to investigate the relationship between HI and star formation in this crucial stellar mass range in which galaxies become less actively star-forming.  We uncover an intriguing population of galaxies at stellar masses log M$_{*}$/M$_{\odot}$ $>$ 10.5 that exhibit surprisingly low sSFRs despite their significant amounts of HI.  Because the galaxies were selected based on single-dish HI observations with Arecibo, there existed no information about the morphology or spatial extent of the HI, which is crucial in determining why much of the HI is not participating in star formation.  In this paper we describe the sample and report on the results of an HI imaging survey at the Jansky Very Large Array (VLA) intended as an observational test of the mechanisms acting to suppress star formation in these massive HI-rich galaxies.

This paper is organized as follows. In Section 2 we select a sample of HI-rich galaxies and we compare their star-forming and structural parameters to two control samples.  
This section motivates the observational test described in Section 3, in which we obtain HI imaging for massive HI-rich galaxies with suppressed star formation.  In Section 4 we discuss the implications of our results in terms of star formation suppression mechanisms.  We summarize our findings in Section 5.  In the Appendix we list notes on individual galaxies.

\section{PROPERTIES OF MASSIVE HI-RICH GALAXIES}
\label{sec:part1}

\subsection{Sample Selection}
\label{sec:part1sample}

GASS \citep{Catinella2010} is a targeted HI survey at Arecibo of $\sim$800 galaxies with stellar masses in the range 10 $<$ log M$_{*}$/M$_{\odot}$ $<$ 11.5 and redshifts in the range 0.025 $<$ z $<$ 0.05.  Each galaxy observed for GASS also lies in the ALFALFA, SDSS, and GALEX footprints, which yield homogeneously measured star formation rates and stellar masses for the sample.  Because GASS is unbiased and complete with respect to stellar mass, we can use it to define scaling relations and select samples based on those scaling relations that are representative of the galaxy population in the local universe.  We use the GASS representative sample from Data Release 2 \citep[N=480;][]{Catinella2012} 
to define a sample of galaxies that are HI-rich for their stellar mass.  (We defined the HI-rich sample and the subsample for follow-up observations before the final data release in \citet{Catinella2013} was available, though that could be used now and should yield the same trends.)  Our simple selection criterion, based only on HI mass and stellar mass, defines the HI-rich sample to contain galaxies that have HI gas fractions in the top 5\% of the GASS distribution.  To determine where the top 5\% lies, we sort the HI gas fractions, or upper limits on the gas fractions for non-detected GASS galaxies, and select the gas fraction that divides the bottom 95\% from the top 5\% (Fig. \ref{fig:gasrich}a) in six evenly spaced stellar mass bins in the range 10.0 $<$ log M$_*$/M$_{\odot}$ $<$ 11.5.  We fit a line to these points that is parametrized as

\begin{equation}
log M_{HI}/M_{\odot} = \alpha (log M_{*}/M_{\odot} -10) + k
\end{equation}

where $\alpha$ = -0.75 and $k$ = 0.02.  We consider all galaxies above the line to be HI-rich for their stellar mass.  A similar sample could be constructed using results of the bivariate HI mass function \citep{Lemonias2013}, but we chose this method for simplicity.

Although we use GASS to establish the criterion for selecting HI-rich galaxies, we select the HI-rich sample from ALFALFA, a blind, shallow wide-field HI survey that contains many more HI detections than GASS and whose sky footprint overlaps with that of GASS.  Whereas GASS provides a complete census of HI in massive galaxies and can be used to derive quantiles of the distribution, ALFALFA mainly detects HI-rich galaxies at the redshifts probed by GASS and so cannot be as easily used to quantify the full range of HI masses for massive galaxies.  However, ALFALFA is complete over this volume to the HI-rich gas fraction limits considered here and so is useful for selecting a large, unbiased sample of HI-rich galaxies.  We select the HI-rich sample from the $\alpha$.40 subsample \citep{Haynes2011}.  To ensure that the physical parameters for galaxies in the HI-rich and GASS samples are measured homogeneously, the HI-rich sample contains only ALFALFA galaxies that are also in the GASS parent sample (N=12006).  There are 1102 unique matches (see Fig. \ref{fig:gasrich}b) between the two samples, of which 258 meet the HI-rich criterion and do not appear to be contaminated by neighboring galaxies.  The final HI-rich sample contains 258 galaxies with total HI masses in the range 10.0 $<$ log M$_{HI}$/M$_{\odot}$ $<$ 10.75 and a median mass of log M$_{HI}$/M$_{\odot}$ $\sim$ 10.26.  The median gas fraction for the sample is 62$\pm$38\%.

To compare to the HI-rich sample, we select a separate sample of ALFALFA galaxies with gas fractions in the range -0.7 $<$ log M$_{HI}$/M$_{*}$ $<$ -0.4 across the full range of stellar mass.  ALFALFA is also complete to these limits over the given redshift volume.  We call this sample the ``constant gas fraction" sample, which contains 285 galaxies.  The median gas fraction for the sample is 29$\pm$6\%.  The lower scatter in the average gas fraction compared to the HI-rich sample is by design.

\subsection{Derived Quantities}
\label{sec:part1data}

The calculation of the SFRs used in this paper is described in detail in \citet{Schiminovich2010}.  We use their DC,D4NCUT SFRs, which are derived directly from near-UV luminosities.  Star-forming galaxies, defined as having a 4000\AA break strength D$_n$(4000) $<$ 1.7, are corrected for internal dust attenuation according to the method outlined in \citet{Johnson2007}.  Their method for determining dust corrections, based on UV and optical fluxes, was empirically calibrated using UV+IR measurements from GALEX and Spitzer.  
We calculate the HI gas fraction as the HI mass divided by the stellar mass, M$_{HI}$/M$_{*}$, such that the HI gas fraction can be greater than one.  Stellar masses and AGN classifications are from the MPA-JHU SDSS DR7 catalog\footnote{http://www.mpa-garching.mpg.de/SDSS/DR7/}.  AGN classes are described in \citet{Kauffmann2003agn}.  
Galaxy sizes (e.g. $R_{90}$, $R_{50}$) and axis ratios are from the NASA-Sloan Atlas\footnote{http://www.nsatlas.org/}.  We define edge-on galaxies as those with axis ratio $\frac{b}{a} <$ 0.3.  
In Section \ref{sec:sflaw} we estimate the SFR and HI surface densities of GASS galaxies by assuming that the star formation is contained within R$_{90}$ and that half of the total HI is contained within R$_{90}$.  The latter is consistent with \citet{Wang2013}, who show that the radius enclosing half of the HI flux (their R50) for normal galaxies is comparable to or less than the optical radius (half of D$_{25}$).

\subsection{Results}
\label{sec:part1results}

\begin{figure*}[h]
\epsscale{0.8}
\plotone{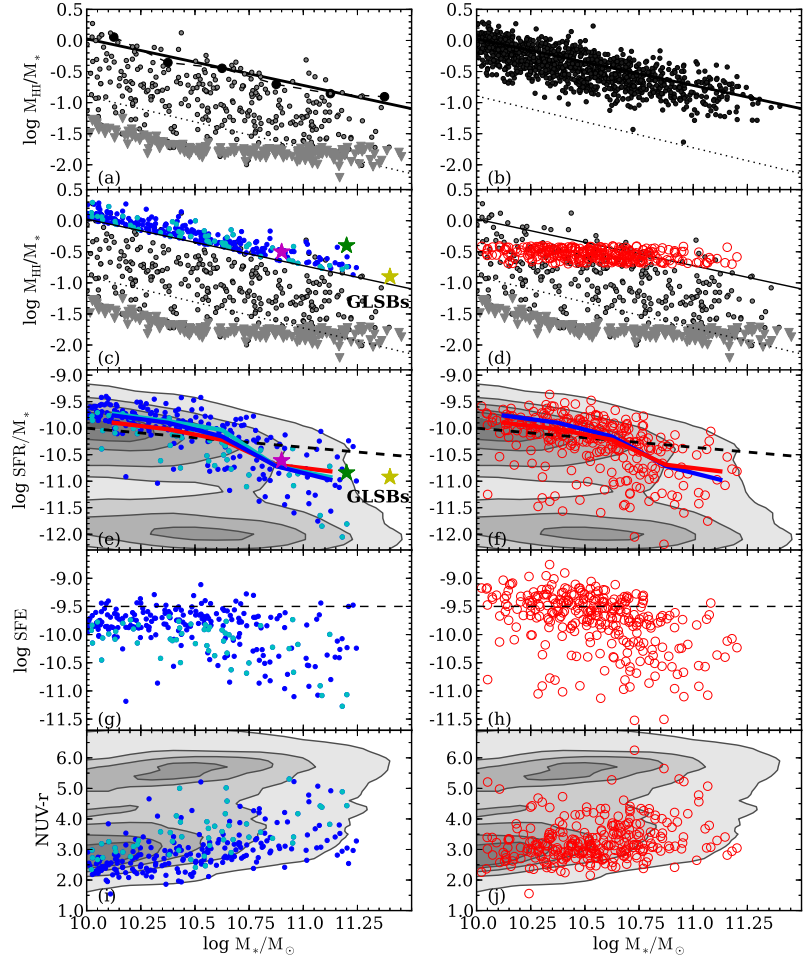}
\caption{The top row shows the selection of the HI-rich galaxies.  The top left panel shows HI gas fraction vs. stellar mass for GASS detections (gray points) and upper limits for GASS non-detections (gray triangles).  The solid line shows the HI-rich selection, and the dotted line is the median of the GASS distribution.  The top right panel shows the ALFALFA detections in the GASS parent sample.  In the bottom four rows we compare the HI-rich sample (left column; blue points) to the constant gas fraction sample (right column; red circles) in terms of HI gas fraction, sSFR, star formation efficiency, and NUV-r color.  Edge-on galaxies are highlighted as cyan points.  Stars indicate the HI fractions and sSFRs of giant low surface brightness galaxies from \citet{Wyder2009}.  In panels $e$ and $f$, blue and red lines show the running medians for the HI-selected samples.  The cyan line shows the same for the HI-rich sample excluding edge-on galaxies.  The black dashed line in panels e and f shows the SF sequence from \citet{Schiminovich2007}.  The black dashed line in panels g and h shows the average SFE from \citet{Schiminovich2010}.  Contours enclose 5\%, 10\%, 25\%, 50\%, and 95\% of the GASS parent sample. } 
\label{fig:gasrich} 
\end{figure*}

First we examine how the SFRs of galaxies in the HI-rich sample vary with stellar mass.  One might naively expect HI-rich galaxies to have uniformly high SFRs.  To test this, in Figs. \ref{fig:gasrich}c,e we plot the gas fractions and sSFRs of the HI-rich galaxies vs. stellar mass.  Two distinct results are apparent: 1) The sSFRs of the HI-rich galaxies decrease with increasing stellar mass faster than their HI contents decrease; and 2) the scatter in their sSFRs increases with stellar mass, leaving many HI-rich galaxies well below the star-forming sequence (dashed line; from \citet{Schiminovich2007}) at log M$_{*}$/M$_{\odot}$ $>$ 10.5.  This population is the subject of our study in the next section.

Our HI-rich sample contains galaxies that have extreme HI contents \emph{for their stellar mass} so their HI fractions necessarily decrease with stellar mass.  But the decreasing gas fraction cannot be the sole driver of the decreasing trend in sSFR because the sSFRs decline with a steeper slope ($\alpha \sim$ -1.29).  To emphasize that the decreasing sSFRs are characteristic of HI-rich galaxies in general and not just a sample of HI-rich galaxies whose HI gas fractions decrease with stellar mass, we compare the HI-rich sample to the constant gas fraction sample, whose galaxies have higher-than-average gas fractions within the same narrow range with respect to stellar mass.  Even though the gas fractions for galaxies in this sample remain constant with respect to stellar mass, their sSFRs decrease in the same way as for the HI-rich sample.  This inconsistent relationship between gas and star formation runs counter to the simple assumption that star formation efficiency (SFE=SFR/M$_{HI}$) does not vary with stellar mass \citep{Schiminovich2010} and hints that something other than the lower gas fractions in massive HI-rich galaxies contributes to their low sSFRs. 

Importantly, this result is not just a reflection of the generic trend of decreasing sSFRs vs. stellar mass \citep[e.g.][]{Schiminovich2007} because we selected HI-rich galaxies which, by definition, are not representative of the full galaxy population. Thus, the decreasing sSFRs within the HI-rich sample is an unexpected result rather than a confirmation of expectations.  That sSFRs decrease with stellar mass even within a sample of galaxies that have extremely high HI contents indicates that gas loss alone cannot explain why sSFRs decrease with stellar mass within the larger population.  By studying this extreme population of galaxies with high HI fractions and incommensurately low SFRs, we can gain insight into the star formation suppression mechanisms at work both in HI-rich galaxies and in less extreme form in normal galaxies.  
Fig. \ref{fig:gasrich}g,h we show the SFE vs. stellar mass for the two HI-selected samples.  The average SFE for GASS galaxies is constant with respect to stellar mass (dashed line) but exhibits a large scatter \citep{Schiminovich2010}.  Overall the HI-rich sample exhibits SFEs lower than the median.  There is also a clear trend for the distribution of SFEs to become wider and extend to lower SFEs at higher stellar masses, which confirms the increasingly weak relationship between HI and star formation within the HI-rich sample.

Finally, in Fig. \ref{fig:gasrich}i,j we show the NUV-r color of the samples, which is generally considered a proxy for specific SFR.  Indeed, the distributions of NUV-r and specific SFR tell a similar story: at low stellar masses, HI-rich galaxies exhibit a narrow range of colors and lie on the blue sequence, but as stellar mass increase, galaxies become redder and the scatter in color increases.  Although the massive HI-rich galaxies have much lower sSFRs than would be expected based on their HI contents, the galaxies do not lie on the red sequence.  They appear to have a low level of residual or recent star formation.

Because the two HI-selected samples have different HI properties by definition but produce similar trends in sSFR, there is some ambiguity in the role of the HI.  It is only clear that within this stellar mass and HI gas fraction range, stellar mass is a better predictor of sSFR than is HI content.  A clearer understanding of these trends requires knowledge of the molecular gas as well, which does not exist for this sample, though we discuss the possible relationship between HI and H$_2$ in this sample below.

We determined that our UV-based SFRs, which can be strongly affected by dust, are not artificially low due to dust attenuation by examining their infrared colors \citep{Wright2010} and D$_{n}$(4000), and removing from the sample edge-on galaxies whose UV fluxes are most likely to be attenuated by gas.  The distribution of D$_{n}$(4000), a stellar age indicator that correlates with star formation history and is expected to be insensitive to extinction, mimics the distribution of sSFR, confirming the existence of a weaker link between gas fraction and sSFR at higher stellar masses.  Although sSFRs for some edge-on galaxies may be underestimated by a factor of ~10, it is unlikely that the decreasing sSFRs with stellar mass are driven by a higher proportion of edge-on galaxies at high stellar masses.  To be sure, we show in Fig. \ref{fig:gasrich} that our results hold if we exclude edge-on galaxies from the HI-rich sample.  Finally, we note that there is generally more scatter in UV-derived SFR measurements for more massive galaxies \citep[see][]{Schiminovich2010}, but it is unlikely to produce the strong trend we see.

The sSFRs of galaxies in the HI-rich sample show that the relationship between HI and star formation varies with stellar mass: at high stellar masses, star formation is suppressed even within galaxies that are HI-rich for their stellar mass.  In the next section we describe HI imaging of 20 HI-rich galaxies with low sSFRs.  The resulting HI maps provide insight into the physical conditions that can suppress star formation while maintaining a significant HI reservoir at high stellar masses.

\section{HI IMAGING OF LOW-STAR-FORMING HI-RICH GALAXIES}
\label{sec:part2}

The low sSFRs in some massive HI-rich galaxies suggests that star formation is being suppressed in some massive galaxies \emph{without the removal of HI}.  The high gas fractions and low sSFRs in the most massive galaxies could be a signature of internal suppression of star formation, extended HI disks with surface densities below that required for efficient star formation, or recently accreted gas that has not yet formed stars.  The distribution and surface density of the HI can provide key insight into the physical conditions of the cold gas in the galaxy, which can help us understand why the cold gas is not forming stars.  We obtained followup HI imaging with the Jansky Very Large Array (VLA) for a subset of the HI-rich galaxies to understand the star formation suppression mechanisms at work in these galaxies.  Below we describe the galaxies selected for followup, the observations, and the conclusions we can derive from the observations.

\subsection{Sample}
\label{sec:part2sample}

\begin{figure*}[h]
\epsscale{0.8}
\plotone{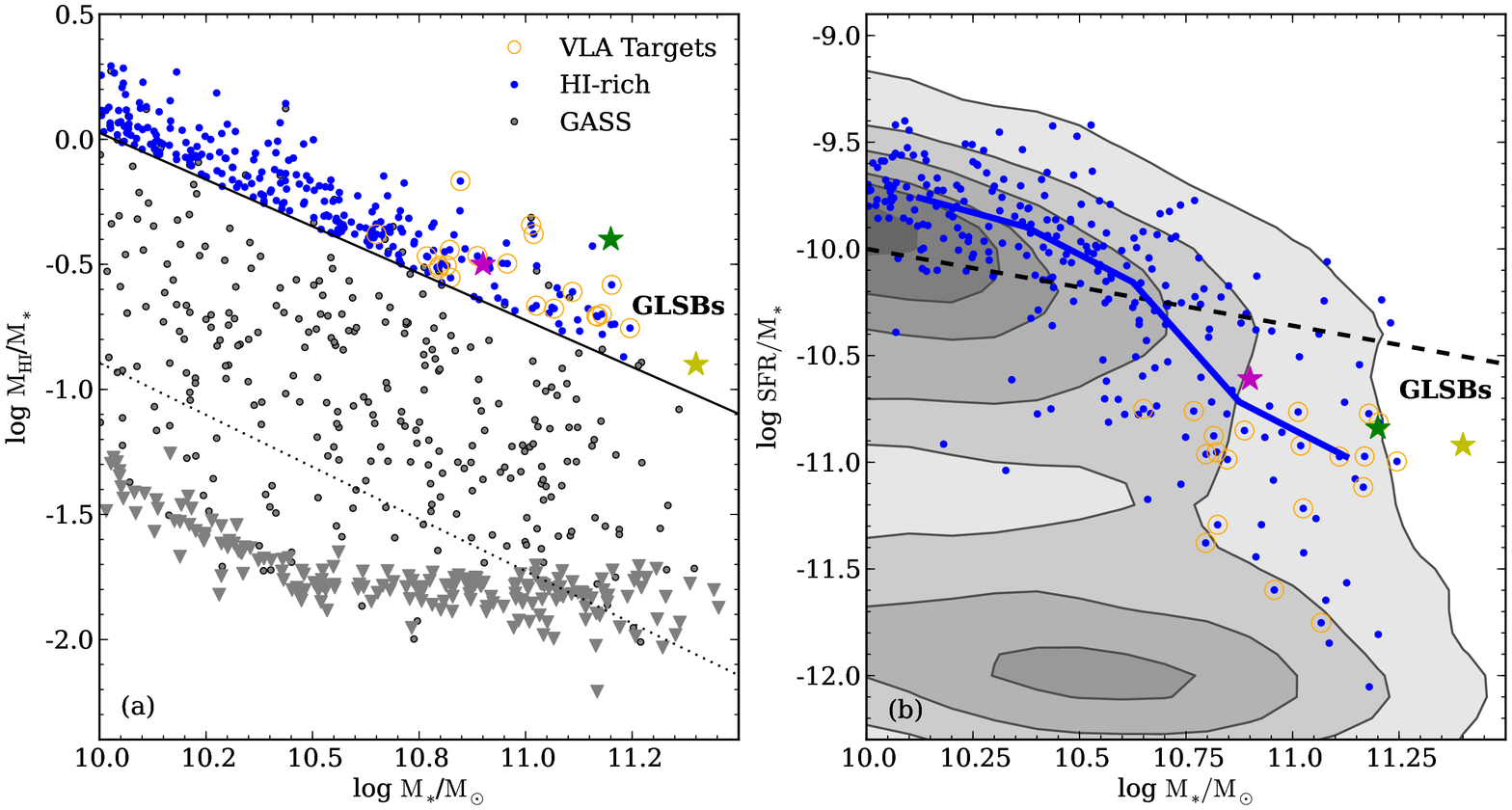}
\caption{Left: HI gas fraction vs. stellar mass for GASS detections (gray points), upper limits for GASS non-detections (gray triangles), and HI-rich sample (blue points).  Solid line shows HI-rich selection.  Dotted line is the median of the GASS distribution.  Stars indicate giant low surface brightness galaxies from \citet{Wyder2009}.  VLA targets are highlighted with black circles.  Right: sSFR vs. stellar mass for HI-rich sample.  Blue line shows the running median.  Black dashed line shows the star-forming sequence from \citet{Schiminovich2007}.  Contours enclose 5\%, 10\%, 25\%, 50\%, and 95\% of the GASS parent sample.} 
\label{fig:vlatargets} 
\end{figure*}

\begin{figure*}[h]
\epsscale{0.8}
\plotone{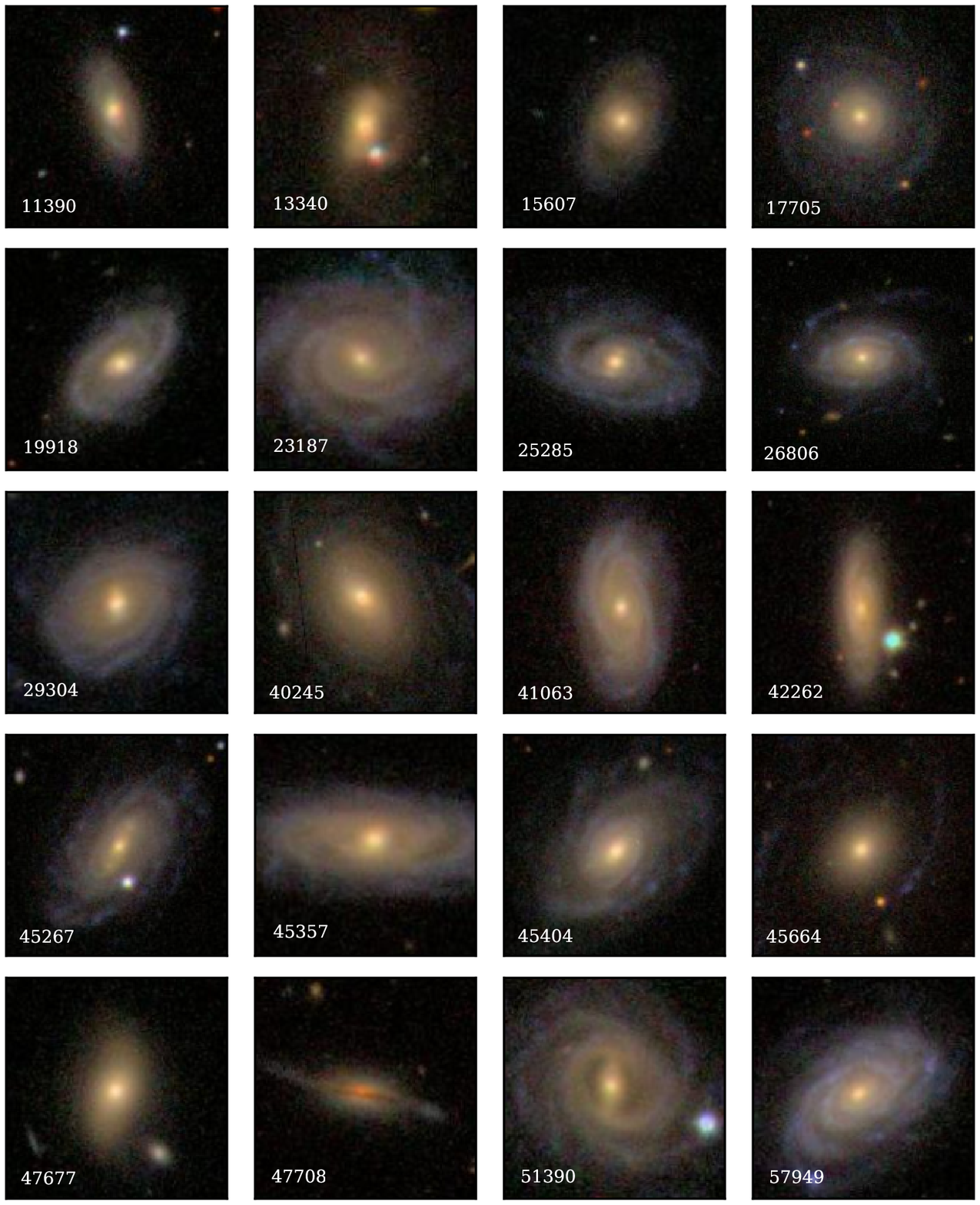}
\caption{50 kpc $\times$ 50 kpc SDSS color images of the 20 galaxies in the VLA sample.} 
\label{fig:sdss} 
\end{figure*}

Galaxies in our HI-rich sample are ideal for followup with the VLA because their high HI fluxes (1.38 - 8.1 Jy km s$^{-1}$), previously measured at Arecibo for the ALFALFA survey, ensure they can be easily detected in a short amount of time.  From the final HI-rich sample of 258 galaxies we selected 20 galaxies for followup with the VLA from the population of galaxies with high HI masses and low sSFRs, i.e. those that deviated from the star-forming sequence at high stellar mass.  The final sample includes all HI-rich galaxies with high stellar masses (log M$_{*}$/M$_{\odot}$ $>$ 10.6), low sSFRs (log sSFR $<$ -10.75), and moderate-to-high axis ratios (b/a $>$ 0.5) to avoid edge-on galaxies whose sSFRs might be artificially low due to internal dust attenuation.  We refer to this subset of HI-rich galaxies as the VLA sample.  They are indicated in Fig. \ref{fig:vlatargets} and their properties are listed in Table \ref{tbl:general}.  In Fig. \ref{fig:sdss} we display their SDSS color images.

One galaxy, GASS 11390, was removed from the sample because the VLA observations revealed no HI signal above the level of the noise.  In the Arecibo spectrum the galaxy is at the edge of the band, which possibly affected the calculation of the HI mass for ALFALFA.  We report on the results of the final sample below, which contains 19 galaxies.  

We note that there exist several other ongoing studies of HI-rich galaxies and we wish to distinguish our sample and our goals from theirs.  The Bluedisks project, as described in \citet{Wang2013} uses HI maps to understand the origin of the excess HI in galaxies that have high HI gas fractions compared to their predicted gas fractions.  Their goal contrasts slightly with our goal, which is to understand why some galaxies with large HI reservoirs do not have high sSFRs.

\citet{Huang2012} describe a study of HI-rich galaxies observed by ALFALFA, HIghMass, that will include a multiwavelength analysis of galaxies with log M$_{HI}$/M$_{\odot}$ $>$ 10.0 and high HI fractions.  Their goal is similar to ours: understanding why some galaxies maintain large gas reservoirs without forming stars at a high rate, but they also include less massive galaxies. 

\citet{Lee2014} present CO observations of a sample of 28 HI-rich galaxies, including 8 LSB galaxies, all of which have relatively high stellar masses (log M$_{\odot}$/M$_*$ $>$ 9.6) and HI masses (log M$_{HI}$/M$_*$ $>$ 10.2).  We refer to this work in Section \ref{sec:H2} when we discuss the possible molecular gas content of galaxies in our sample.

\begin{deluxetable*}{cccccccccccc}
\tabletypesize{\scriptsize}
\tablecolumns{12}
\tablecaption{General Data}
\tablewidth{0pt}
\tablehead{
\colhead{GASS ID} & \colhead{RA} & \colhead{DEC} & \colhead{z} & \colhead{log M$_{*}$} & \colhead{log M$_{HI}$} &
\colhead{SFR} & \colhead{sSFR} & \colhead{NUV-r} & \colhead{R$_{90}$/R$_{50}$} & \colhead{b/a} & \colhead{AGN Class} \\
\tableline\\
\colhead{} & \colhead{} & \colhead{} & \colhead{} & \colhead{M$_{\odot}$} & \colhead{M$_{\odot}$} & \colhead{M$_{\odot}$ yr$^{-1}$} & \colhead{yr$^{-1}$} & \colhead{mag.} & \colhead{} & \colhead{} & \colhead{}
}
\startdata
11390	&350.86826	&13.97744	&0.04130	&10.8	&10.3	&0.3	&-11.3	&4.0	&3.2	&0.6	&Weak AGN\\
13340	&198.38022	&5.91552	&0.04859	&11.0	&10.5	&0.2	&-11.6	&5.2	&2.8	&0.5	&Composite\\
15607	&168.10269	&7.49480	&0.04721	&10.8	&10.3	&0.3	&-11.4	&4.2	&3.2	&0.9	&Weak AGN\\
17705	&164.93533	&10.07098	&0.03545	&10.9	&10.4	&1.1	&-10.9	&3.1	&2.9	&0.9	&Low S/N AGN\\
19918	&129.69577	&7.48071	&0.04624	&10.8	&10.4	&0.7	&-11.0	&3.5	&2.6	&0.7	&Low S/N AGN\\
23187	&158.47237	&11.20706	&0.04984	&11.2	&10.6	&2.4	&-10.8	&3.0	&1.9	&0.7	&...\\
25285	&200.38412	&12.18775	&0.03812	&10.8	&10.3	&1.0	&-10.8	&3.0	&2.6	&0.8	&Low S/N AGN\\
26806	&133.29718	&9.14816	&0.02924	&10.6	&10.3	&0.8	&-10.8	&3.2	&2.7	&0.7	&Low S/N AGN\\
29304	&221.58905	&13.02097	&0.04672	&11.0	&10.7	&1.8	&-10.8	&3.0	&2.4	&0.8	&...\\
40245	&209.85847	&12.78848	&0.03906	&11.2	&10.5	&1.1	&-11.1	&3.8	&3.2	&0.7	&...\\
41063	&207.34860	&8.50763	&0.03790	&11.1	&10.5	&1.4	&-11.0	&3.3	&2.1	&0.6	&...\\
42262	&236.97530	&25.72956	&0.04183	&11.0	&10.4	&0.6	&-11.2	&4.8	&2.6	&0.4	&Low S/N AGN\\
45267	&214.59982	&26.38588	&0.03625	&10.8	&10.3	&0.9	&-10.9	&3.1	&2.5	&0.5	&Low S/N AGN\\
45357	&215.38055	&23.94831	&0.04993	&11.2	&10.5	&1.8	&-11.0	&3.3	&2.6	&0.5	&...\\
45404	&213.12100	&24.63550	&0.03803	&11.0	&10.6	&1.2	&-10.9	&3.5	&3.2	&0.6	&Low S/N AGN\\
45664	&216.19432	&26.13968	&0.03642	&10.8	&10.7	&0.7	&-11.0	&3.4	&3.1	&0.8	&Weak AGN\\
47677	&167.75920	&25.90686	&0.04073	&11.1	&10.4	&0.2	&-11.8	&5.1	&3.2	&0.7	&Weak AGN\\
47708	&165.58934	&25.66284	&0.04494	&10.8	&10.3	&0.7	&-11.0	&4.3	&2.4	&0.5	&Low S/N AGN\\
51390	&119.06930	&11.66217	&0.04566	&11.2	&10.5	&1.6	&-11.0	&3.5	&2.0	&0.7	&Low S/N AGN\\
57949	&178.29640	&25.43762	&0.04437	&11.2	&10.5	&2.6	&-10.8	&2.9	&2.2	&0.6	&...
\enddata
\label{tbl:general}
\end{deluxetable*}

\subsection{Observations and Data Reduction}
\label{sec:part2obs}

\begin{figure*}[t]
\epsscale{1.0}
\plotone{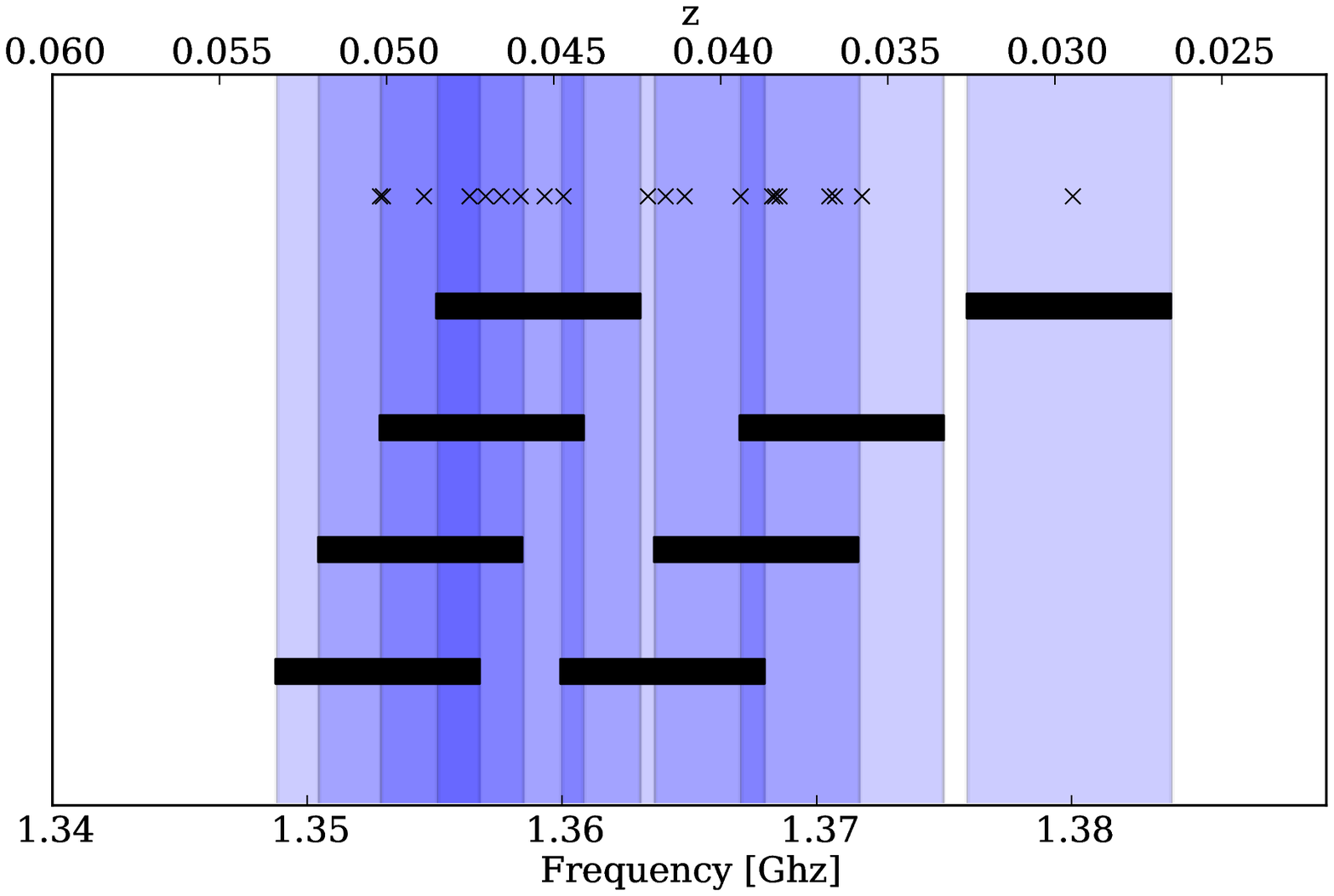}
\caption{Instrument Setup.  Blue bands and black stripes denote the eight spectral windows designed for HI lines.  Each $x$ denotes the redshifted frequency of the HI line for each galaxy.}
\label{fig:instrument} 
\end{figure*}

Observations with the VLA were conducted in June-July 2013 in spectral-line mode at 21 cm in C configuration.  We selected integration times of 2 or 4 hours, including time for calibration (with standard flux calibrators 3C48, 3C147, and 3C286), for each galaxy depending on the HI fluxes detected at Arecibo, optical sizes, and an assumed HI diameter, D$_{HI}$=1.5D$_{opt}$, where D$_{opt}=2R_{90}$ using r-band measurements.  We defined eight spectral windows to cover the velocity range needed to detect HI emission at the systemic velocities of the galaxies (see Fig. \ref{fig:instrument}).  Each spectral window had 256 channels covering a frequency range of 8 MHz for a frequency resolution of 31.25 KHz or a velocity resolution of about 7 km/s. The spectral windows were not evenly spaced in frequency space but were defined so that HI emission from each galaxy would fall close to the middle of a spectral window.  Some of the spectral windows overlapped significantly.  For data reduction we used only the data from the spectral window within which the galaxy was closest to the center (except for GASS40245; see Appendix).

Data were reduced with CASA \citep{CASA2007} following standard calibration procedures.  Bad data points were selected and flagged based on the calibration solutions and by-eye inspection of the visibilities.  We determined the continuum by fitting a line to the line-free channels and subtracted the result.

We used the CASA task CLEAN to create data cubes from the calibrated data.  Data cubes were built with 6$\times$6 arcsecond pixels and a velocity resolution of about 28 km/s by averaging four adjacent channels.  The median size of the synthesized beam is 21$\times$17 arcseconds.  We used a weighting scheme with a robustness parameter of 1 \citep{Briggs1995} to emphasize low signal-to-noise emission.  The typical noise is 0.54 mJy per beam per channel; we CLEANed down to two times the noise level.

From the data cubes, we chose by eye which channels to include in the total HI intensity (moment-0) and velocity (moment-1) maps.  We selected only those channels containing emission that appeared in more than one adjacent channel and seemed to be associated with the galaxy.  For all galaxies, we first chose which pixels to include by adjusting the flux thresholds until we reached a compromise between including enough low surface brightness flux and not too much noise.  In practice, we convolved the original data cube with a 20$\times$20 arcsecond kernel and selected the flux threshold based on this convolved image to avoid including noise in the final moment maps.  This image was used to define which 6$\times$6 arcsecond pixels of the original cube should be included in the moment maps.  Thus, the final moment maps have 6$\times$6 arcsecond pixels based on the original, unsmoothed data cube.  We masked any remaining noise in the moment maps to minimize its effect on the calculation of HI masses, radial profiles, and HI radii.

For three galaxies with low surface brightness HI emission (15607, 19918, 47708) we generated moment maps using a method that is less likely to exclude low levels of emission.  Instead of imposing a threshold, we selected regions of emission in the data cube that appeared in more than one adjacent channel and summed the flux in those regions to produce the HI intensity and velocity maps.

\subsection{Derived Quantities}

Following \citet{Walter2008}, we derived HI quantities directly from the moment maps that were constructed and masked to limit spurious HI.

\emph{HI mass}  We calculated HI masses by summing the flux in the HI intensity maps and converting the flux to HI mass in solar masses according to the following equation:

\begin{equation}
\frac{M_{HI}}{M_{\odot}} = \frac{2.356\times10^5}{1 + z} d_L^2 S
\end{equation}

where d$_L$ is the luminosity distance in Mpc and $S$ is the flux in Jy km s$^{-1}$.

\emph{HI Radius}  We constructed radial profiles by summing the flux in concentric elliptical annuli defined by the axis ratio and position angle of the optical disk reported in the NASA-Sloan Atlas.  The annuli were 6 arcseconds in width; an average of 15 independent data points contributed to the measurement of the HI radii.  HI fluxes were converted to surface brightness in solar masses per square parsec.  We determined the radial profiles out to a radius determined by eye for each galaxy.  We use the radial profiles to calculate R90$_{HI}$, the radius that contains 90\% of the total HI flux.  In the HI intensity maps we also display the contour at which the HI surface brightness drops to 1 M$_{\odot}$ pc$^{-2}$, which is equivalent to a column density of 1.25 $\times$ 10$^{20}$ atoms cm$^{-2}$.  Other methods of calculating the extent of cold gas include measuring the maximum distance to a given isophote \citep{Serra2012, Davis2013}.

\emph{HI surface density}  The surface density of gas can be measured locally, for individual regions within a galaxy \citep[as in, e.g.,][]{Bigiel2008, Bigiel2010, Leroy2008}, and globally, over the entire extent of a galaxy \citep[as in][]{Kennicutt1998}.  To be consistent with the analysis of \citet{Kennicutt1998}, and because the distances to galaxies in this paper are large enough that there are only a few resolution elements per galaxy, we only calculate global HI surface densities.  The global surface density necessarily averages over important variations in the local surface densities; the HI intensity maps in Fig. \ref{fig:mm1} give some indication of these variations. 

There are several ways of calculating the global HI surface density.  To provide a sense of the possible range of global HI surface densities in these galaxies, we report the HI surface density calculated in three ways.  First we calculate the HI surface density averaged over the regions of the galaxy within the 1 M$_{\odot}$ pc$^{-2}$ isophote.  This contour is shown on the HI intensity maps.  A benefit to this method of calculating HI  surface density is that it only includes regions of the galaxy that contain significant amounts of HI.  Second, we use the HI radii derived from these observations, R90$_{HI}$, and the optical radii reported in the NASA-Sloan Atlas, R90$_{opt}$, to compute the HI surface densities within these regions - the HI disk and the stellar disk.  \citet{Kennicutt1998} calculated the global HI surface density within the stellar disk, but this measurement provides no information about the outer parts of the HI disk since HI disks tend to extend well beyond the optical disks of galaxies.

\emph{SFR Surface Density} We also calculate the SFR surface density in three ways, averaging over the same three regions we used for the HI surface density.  To determine the SFR within each region, we scaled the SFR from GALEX by the fraction of far-UV light contained within the given region, which assumes that the dust correction is uniform across the galaxy.

\begin{figure*}[t]
\epsscale{0.9}
\plotone{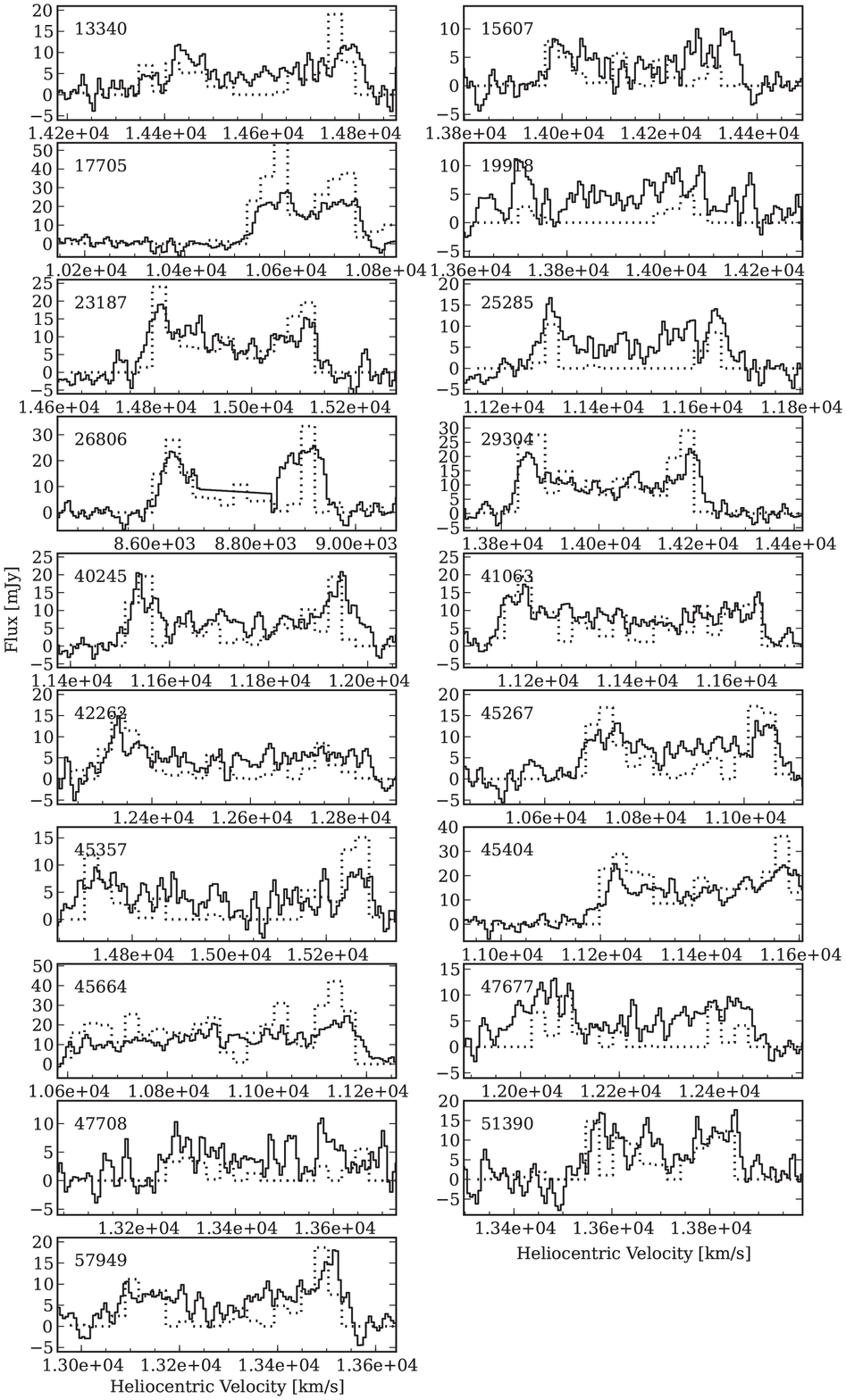}
\caption{Velocity spectra from ALFALFA (solid lines) and this survey (dotted lines).}
\label{fig:velocity} 
\end{figure*}

\emph{Velocity Spectra} In Fig. \ref{fig:velocity} we compare the velocity spectra from this survey to the ALFALFA spectra.  The velocity resolution for the VLA spectra is coarser because we smoothed the cubes as described above.  There is general agreement between the spectra except in cases where we do not recover the total HI mass that ALFALFA detected (see next section).  In some cases the spectra match at the peaks but the VLA is missing some flux close to the systemic velocities.  This flux might be harder to detect because the HI in channels close to the systemic velocity could, in a given channel, cover a larger physical area and subsequently have lower surface brightness below the noise level.  We are probably not missing any high surface brightness gas, but there might be lower surface brightness gas that we are unable to detect, the presence of which would only strengthen the conclusions of this paper.

\subsection{Results}
\label{sec:part2results}

\subsubsection{HI Morphology}

\begin{figure*}[t]
\epsscale{0.8}
\plotone{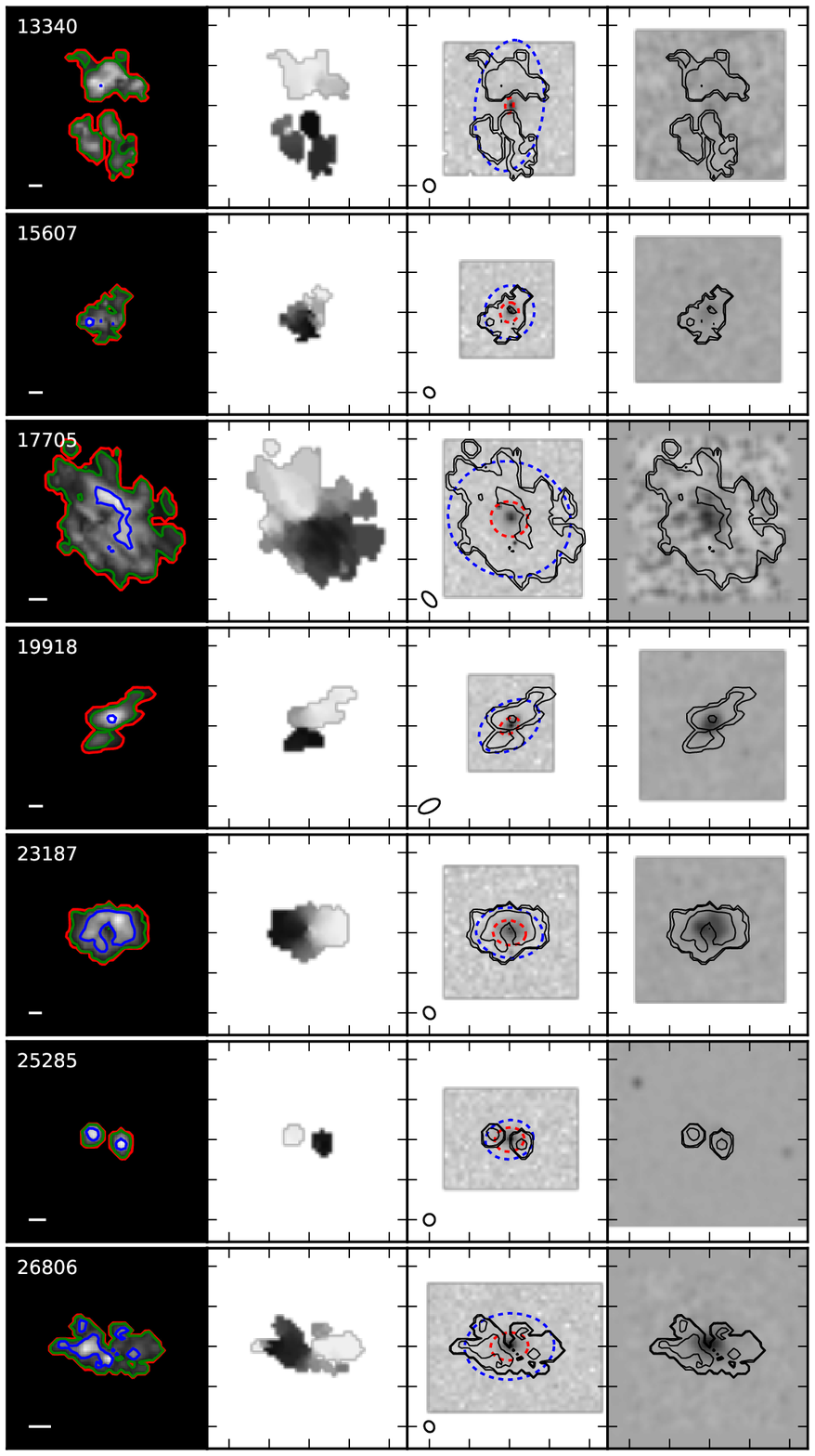}
\caption{HI intensity and velocity maps for galaxies in the VLA sample, 5 arcminutes on a side.  The column on the far left shows the HI intensity maps with lines of constant surface brightness indicated (red, green, and blue contours indicate 0.2, 1.0, and 5.0 M$_{\odot}$ pc$^{-2}$). The scalebar indicates 20 kpc.  The next column shows the velocity maps; the velocity range for each map roughly corresponds to the velocity ranges in Fig. \ref{fig:velocity}.  The third and fourth columns show the surface brightness contours overlaid on the SDSS r-band and GALEX NUV images.  Blue dashed ellipses indicate R90$_{HI}$ and red dashed ellipses indicate R90$_{opt}$.  The beam size is shown in the lower left corner.  SDSS and GALEX images were scaled to the same resolution as the HI maps and were obtained from the NASA-Sloan Atlas and MAST.  }
\label{fig:mm1} 
\end{figure*}

\begin{figure*}[t]
\epsscale{0.8}
\plotone{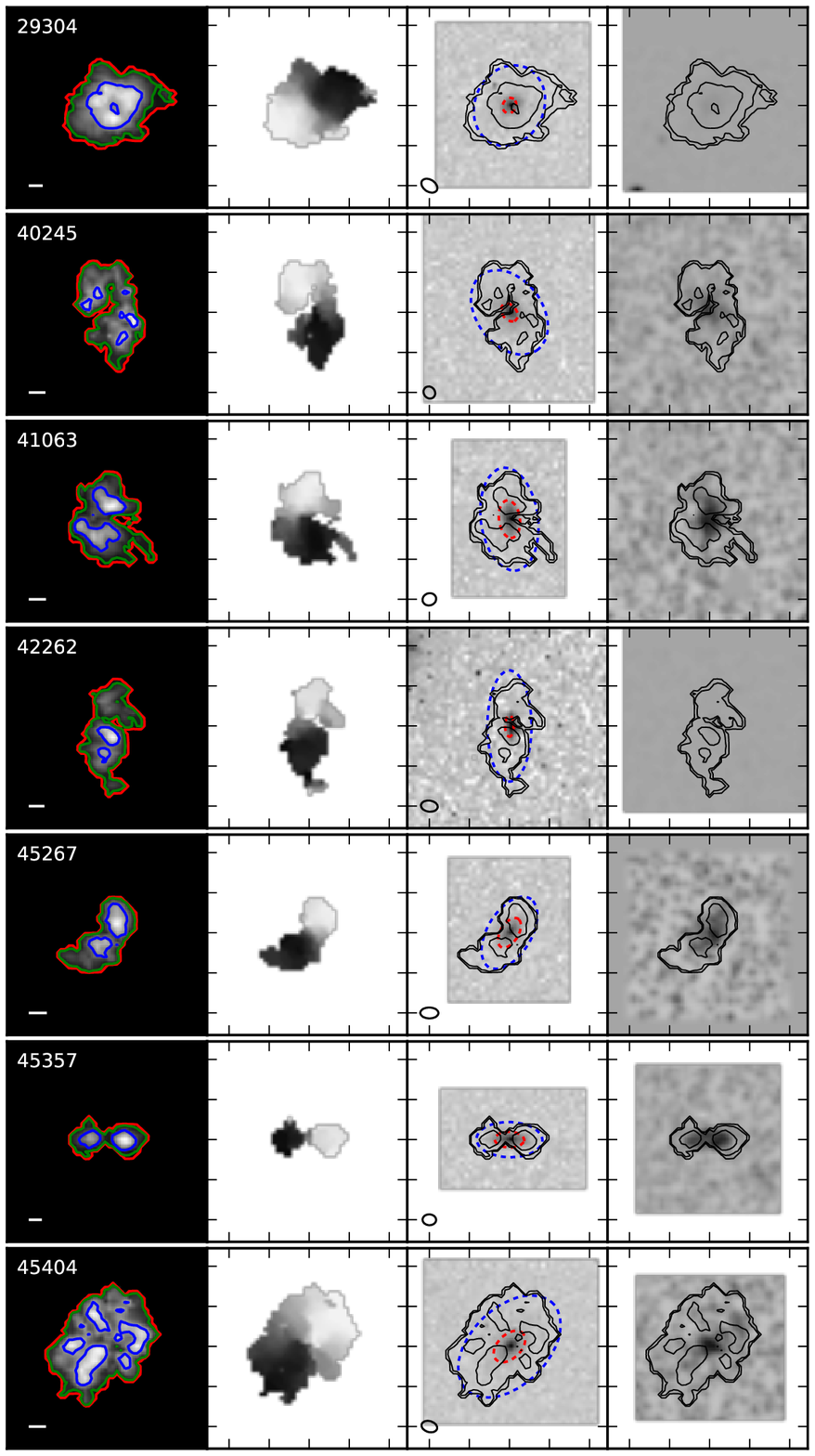}
\caption{See caption for Fig. \ref{fig:mm1}.}
\label{fig:mm2} 
\end{figure*}

\begin{figure*}[t]
\epsscale{1.0}
\plotone{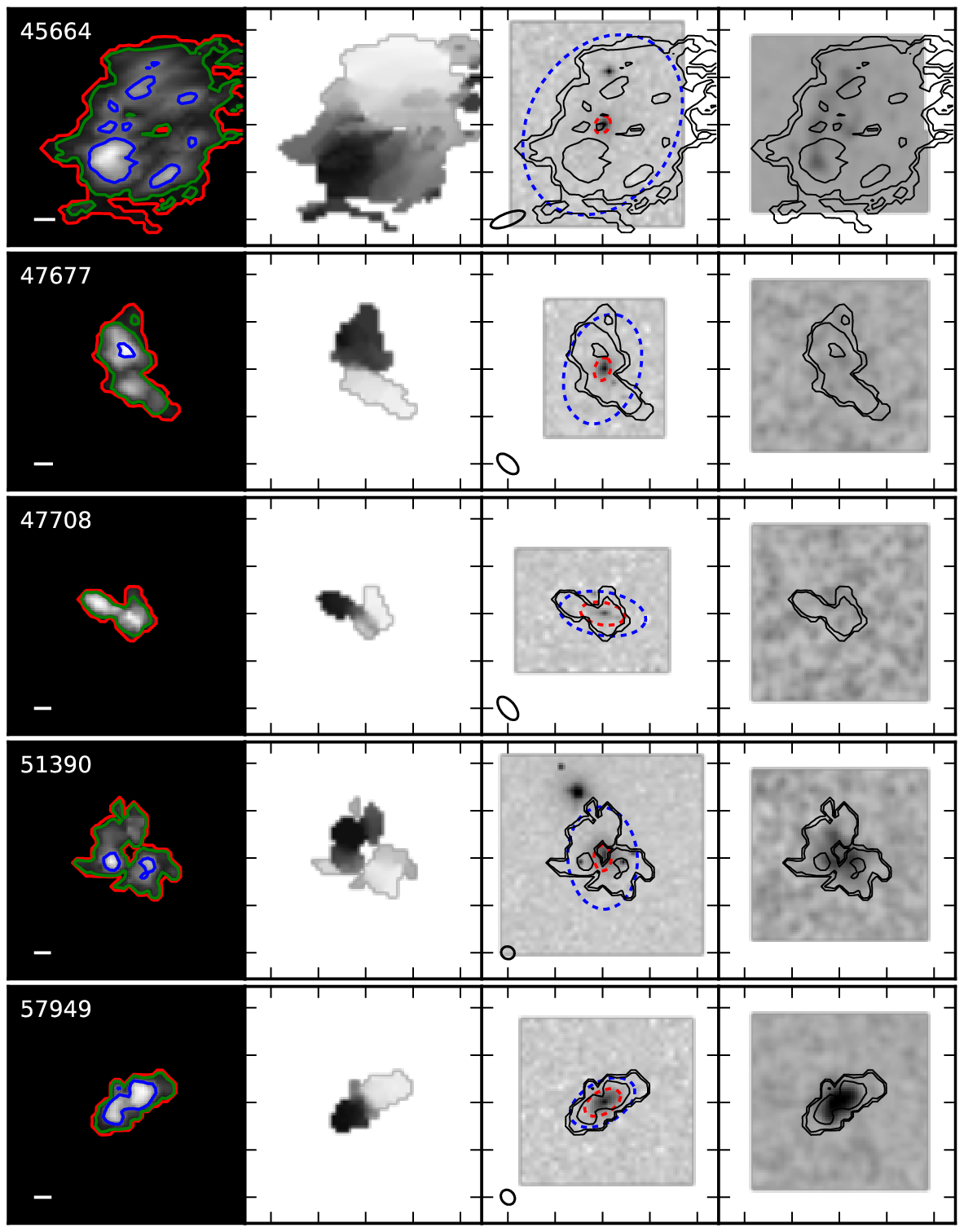}
\caption{See caption for Fig. \ref{fig:mm1}.}
\label{fig:mm3} 
\end{figure*}

In Fig. \ref{fig:mm1} we display the total intensity and velocity maps and the intensity maps overlaid on SDSS and GALEX images.  It is clear that while all of the galaxies in the VLA sample exhibit HI emission well beyond the optical disks, the precise morphology of the HI varies widely.  However, all 19 of our galaxies have HI that extends beyond the stellar disk and exhibit regularly rotating disks, placing them in the \emph{D} category of \citet{Serra2012}.  As pointed out in \citet{Serra2012}, regularly rotating disks of HI have likely been in place for several gigayears, while less settled HI morphologies might be the result of recent accretion or interactions.  We comment on the role of external events such as these below. 

We also calculated the concentration of the HI using the proxy R90$_{HI}$/R50$_{HI}$ and found no clear trends with respect to sSFR or SFE.  This could mean that the star formation suppression mechanisms acting on the gas are independent of the precise distribution of HI.

\begin{deluxetable*}{rrccccccccc}
\tabletypesize{\scriptsize}
\tablecolumns{11}
\tablecaption{Derived HI Quantities}
\tablewidth{0pt}
\tablehead{
\colhead{GASS ID} & \colhead{AA ID} & \colhead{log AA MHI} & \colhead{log VLA MHI} & \colhead{Flux} & \colhead{R$_{90HI}$} & \colhead{R$_{90opt}$} &
\colhead{Vhelio} & \colhead{Frequency} & \colhead{t$_{int}$} & \colhead{Noise} \\
\tableline\\
\colhead{} & \colhead{} & \colhead{M$_{\odot}$} & \colhead{M$_{\odot}$} & \colhead{Jy km s$^{-1}$} & \colhead{kpc} & \colhead{kpc} & \colhead{km s$^{-1}$} & \colhead{MHz} & \colhead{hours} & \colhead{mJy/beam/channel}
}
\startdata
11390	&332599	&10.3	&---	&---	&---	&11.7	&12420	&1361.743	&2	&0.500\\
13340	&232760	&10.5	&10.3	&1.9	&100.0	&12.0	&14608	&1351.388	&4	&0.400\\
15607	&213794	&10.3	&9.9	&0.9	&39.9	&14.8	&14163	&1353.348	&2	&0.535\\
17705	&6072	&10.4	&10.5	&6.4	&68.7	&20.8	&10646	&1370.052	&2	&0.559\\
19918	&181103	&10.4	&9.9	&0.8	&48.3	&15.5	&13933	&1354.726	&2	&0.524\\
23187	&5737	&10.6	&10.5	&3.4	&52.5	&25.8	&14956	&1349.613	&4	&0.380\\
25285	&8395	&10.3	&9.6	&0.7	&28.6	&17.5	&11462	&1366.260	&4	&0.590\\
26806	&4652	&10.3	&10.1	&3.4	&40.5	&17.0	& 8773	&1378.873	&2	&0.760\\
29304	&9515	&10.7	&10.7	&4.9	&60.6	&23.7	&14017	&1354.044	&2	&0.490\\
40245	&8907	&10.5	&10.2	&2.7	&57.0	&26.0	&11745	&1364.925	&2	&0.510\\
41063	&231506	&10.5	&10.4	&3.8	&61.1	&19.4	&11386	&1366.572	&2	&0.542\\
42262	&251252	&10.4	&10.2	&2.0	&72.5	&13.0	&12575	&1360.990	&2	&0.497\\
45267	&9157	&10.3	&10.1	&2.5	&45.1	&18.1	&10868	&1368.916	&2	&0.700\\
45357	&9195	&10.5	&10.3	&1.9	&51.7	&25.2	&14986	&1349.485	&4	&0.440\\
45404	&9094	&10.6	&10.6	&6.7	&72.4	&19.3	&11410	&1366.388	&2	&0.550\\
45664	&9234	&10.7	&10.8	&10.5	&91.6	&12.8	&10890	&1368.675	&2	&0.740\\
47677	&723039	&10.4	&10.0	&1.6	&60.4	&12.3	&12219	&1362.553	&2	&0.560\\
47708	&722779	&10.3	&9.7	&0.7	&51.7	&20.2	&13483	&1356.573	&2	&0.535\\
51390	&4109	&10.5	&10.3	&2.1	&61.9	&24.8	&13706	&1355.550	&4	&0.500\\
57949	&6861	&10.5	&10.2	&2.1	&41.3	&21.3	&13303	&1357.382	&2	&0.500
\enddata
\label{tbl:HI}
\end{deluxetable*}

\begin{deluxetable*}{rcccccc}
\tabletypesize{\scriptsize}
\tablecolumns{7}
\tablecaption{HI and SFR Surface Densities\tablenotemark{a}}
\tablewidth{0pt}
\tablehead{
\colhead{GASS ID} & \colhead{$\Sigma_{HI,RHI}$} & \colhead{$\Sigma_{HI,Ropt}$} & \colhead{$\Sigma_{HI,1M\odot}$} & \colhead{$\Sigma_{SFR,RHI}$} & \colhead{$\Sigma_{SFR,Ropt}$} & \colhead{$\Sigma_{SFR,1M\odot}$} \\
\tableline\\
\colhead{} & \colhead{M$_{\odot}$ pc$^{-2}$}  & \colhead{M$_{\odot}$ pc$^{-2}$}  & \colhead{M$_{\odot}$ pc$^{-2}$} & \colhead{M$_{\odot}$ kpc$^{-2}$} & \colhead{M$_{\odot}$ kpc$^{-2}$} & \colhead{M$_{\odot}$ kpc$^{-2}$}
}
\startdata
13340	&-0.26	&0.02	&-0.83	&-5.49	&-5.62	&-5.36\\
15607	&0.19	&0.39	&0.44	&-4.28	&-4.12	&-3.57\\
17705	&0.32	&0.41	&0.75	&-4.14	&-4.08	&-3.16\\
19918	&-0.03	&0.20	&0.42	&-4.10	&-3.94	&-3.15\\
23187	&0.58	&0.55	&0.70	&-3.58	&-3.67	&-3.13\\
25285	&0.17	&0.49	&0.39	&-3.50	&-3.50	&-3.15\\
26806	&0.34	&0.51	&0.59	&-3.81	&-3.72	&-3.13\\
29304	&0.56	&0.54	&0.86	&-4.00	&-4.01	&-3.38\\
40245	&0.20	&0.32	&0.20	&-3.99	&-4.03	&-3.58\\
41063	&0.26	&0.36	&0.41	&-4.02	&-4.01	&-3.18\\
42262	&-0.03	&0.09	&-0.07	&-4.51	&-4.86	&-3.54\\
45267	&0.31	&0.35	&0.44	&-3.87	&-3.95	&-3.14\\
45357	&0.34	&0.43	&0.39	&-3.76	&-3.72	&-3.22\\
45404	&0.36	&0.39	&0.43	&-4.12	&-4.13	&-3.30\\
45664	&0.31	&0.41	&0.62	&-4.57	&-4.51	&-3.88\\
47677	&-0.05	&0.26	&0.23	&-4.95	&-4.69	&-4.29\\
47708	&-0.20	&0.10	&0.03	&-4.16	&-3.95	&-3.75\\
51390	&0.16	&0.30	&0.21	&-4.07	&-4.07	&-3.52\\
57949	&0.48	&0.46	&0.64	&-3.42	&-3.49	&-2.89
\enddata
\label{tbl:sigmas}
\tablenotetext{a}{All surface densities are shown in log base 10.}
\end{deluxetable*}

\subsubsection{HI Radii and Masses}

The derived quantities described above are listed in Tables \ref{tbl:HI} and \ref{tbl:sigmas}.  We can compare the HI masses calculated from the VLA observations to the HI masses derived from the ALFALFA Arecibo observations as a check on the VLA observations.  In general, the HI masses derived from the VLA are within 0.3 dex of the ALFALFA masses.  There are four galaxies whose VLA mass deviates from the ALFALFA mass by more than 0.3 dex.  These include 15607, 19918, and 47708, all of which have lower column density HI than most of the other galaxies in the sample.  It is likely that the rest of the HI exists in an extended disk at even lower column densities that the VLA, with our several-hour integration times, was not sensitive to.  If this is the case, then the HI radii should be considered lower limits and the HI surface densities upper limits.  We found no evidence of radio continuum sources that could be absorbing the HI, artificially lowering the HI masses.  Another galaxy with a low HI mass is 25285.  Half of the data for this observation were corrupted and unusable so we didn't achieve the sensitivity we wanted.  
The HI measurements for 25285 should also be considered lower limits.  

We take the systematic and statistical errors on the HI masses as the lower and upper bounds on the errors.  We take the systematic error to be 0.18 dex, the mean discrepancy between the ALFALFA and VLA masses.  We calculate the statistical error on the HI mass to be 0.04 dex based on the average noise per channel of 0.5 mJy/beam.  The error on the HI radius will scale with this error since our ability to calculate accurately the HI radius depends on our ability to detect HI.

\begin{figure*}[t]
\epsscale{1.0}
\plotone{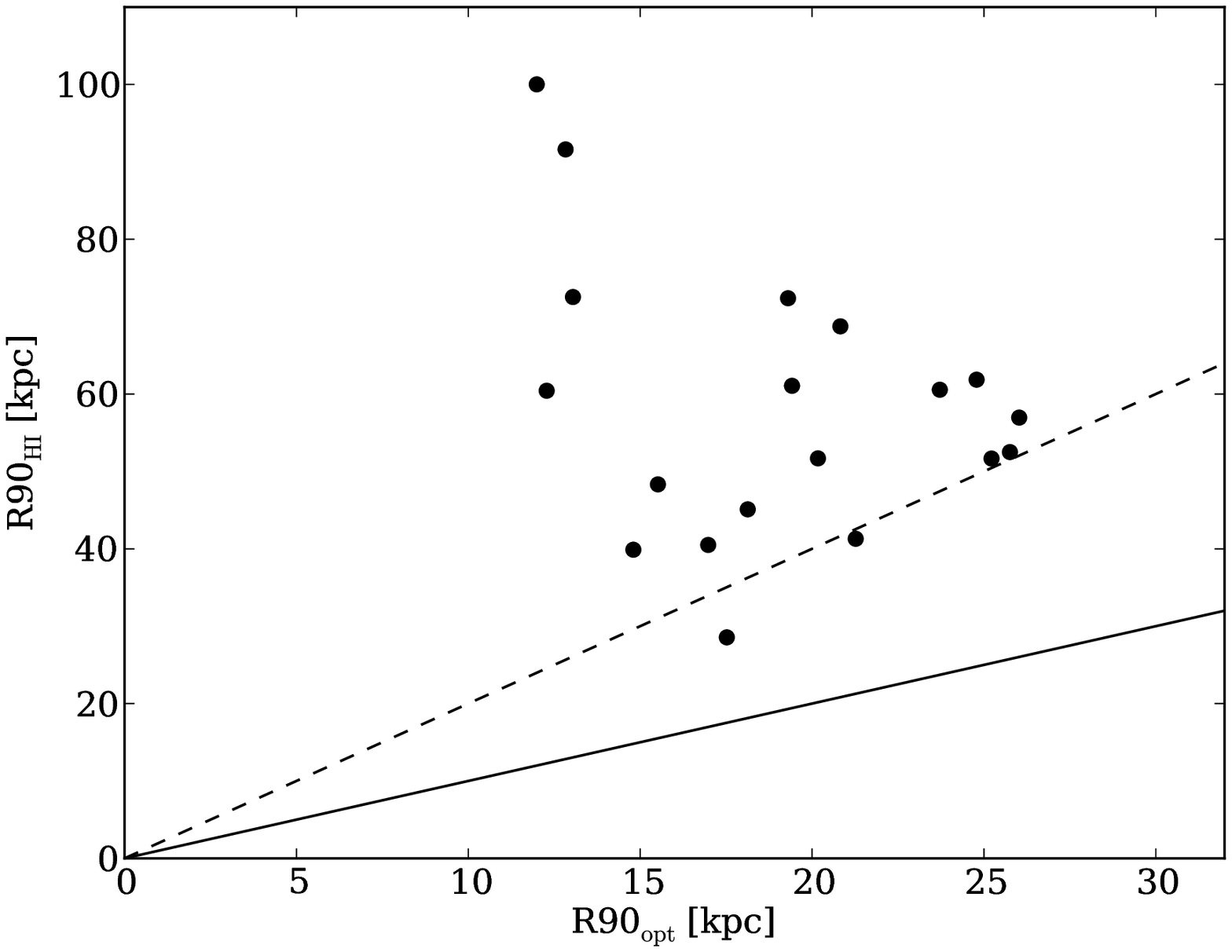}
\caption{R90$_{HI}$ vs. R90$_{opt}$.  Solid line shows 1-to-1 relation and dashed line shows 2-to-1 relationship.  Most galaxies have HI radii at least twice as large as their optical radii.}
\label{fig:HIradius} 
\end{figure*}

In Fig. \ref{fig:HIradius} we compare R90$_{HI}$ to R90$_{opt}$.  All of the galaxies in our sample have HI disks that extend beyond the optical disks and almost all of the HI disks have radii that are twice as large as the optical radii.  The median ratio of R90$_{HI}$/R90$_{opt}$ is 2.6.  There exists no obvious comparison sample since galaxies in the VLA sample span a wide range of morphologies and a narrow range of stellar masses and HI masses.   The HI-to-optical ratio for our sample is higher than that for a sample of 68 early-type disk galaxies (S0-Sab) in \citet{Noordermeer2005}, who report ratios of 1.72 for Sa/Sab galaxies and 2.11 for S0/S0a galaxies.  While only 2 out of 68 galaxies in their sample have R$_{HI}$ $>$ 40 kpc, almost all of the galaxies in our sample have R$_{HI}$ $>$40 kpc.  A noticeable difference between their sample and ours is that 18\% of the galaxies in their sample have HI disks that lie within the stellar disks.  They ascribe the smaller HI disks to ram-pressure stripping and other types of interactions because most of the galaxies with R$_{HI}$ $<$ R$_{opt}$ show signs of interactions.  Our results here are consistent with that found in \citet{Wang2013} for a sample of HI-rich galaxies, though their sample includes some galaxies with lower stellar masses and lower HI masses.

\subsubsection{The Star Formation Law}
\label{sec:sflaw}

To understand the origin of the low sSFRs for galaxies in the VLA sample, it is crucial to determine not just the extent of their HI disks but also the surface density of HI within the HI disks.  The HI surface density provides information about which types of suppression mechanisms are at work in these galaxies.  Based on work showing that star formation is inefficient at low gas surface densities, low HI surface densities alone could explain low sSFRs.  If galaxies in the VLA sample have high HI surface densities, then something else must be preventing the gas from forming stars.  In combination with the SFR surface density, the HI surface density can tell us if these galaxies follow well-established relationships between cold gas and star formation, with both low gas and SFR surface densities as in the lower part of the star formation law, or if there is something unique about these galaxies.

\begin{figure*}[t]
\epsscale{1.0}
\plotone{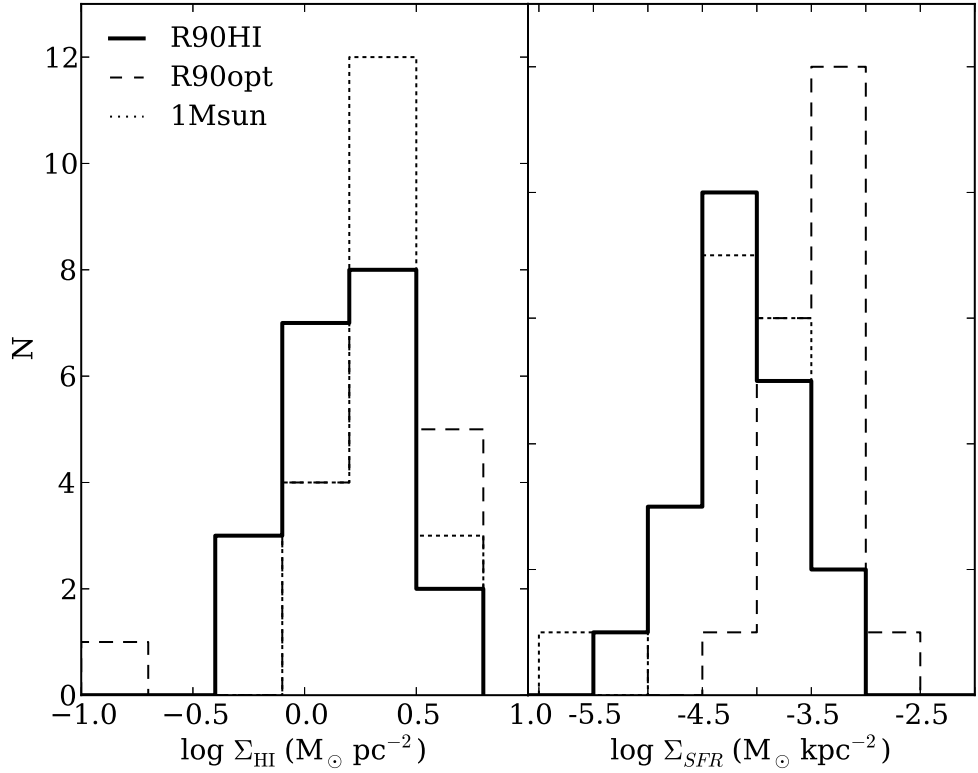}
\caption{Histograms of the three measurements for HI surface density (left panel) and SFR surface density (right panel).  Surface densities within the HI disk are indicated by solid lines, within the stellar disk by dashed lines, and within the 1 M$_{\odot}$ pc$^{-2}$ isophote by dotted lines.}
\label{fig:sigmaHI} 
\end{figure*}

We compare the three calculations of HI surface densities in the left panel of Fig. \ref{fig:sigmaHI}.  The HI surface density calculated within the 1 M$_{\odot}$ pc$^{-2}$ isophote and within R90$_{HI}$ have similar distributions.  Since Fig. \ref{fig:mm1} shows that R90$_{HI}$ and the 1 M$_{\odot}$ pc$^{-2}$ isophote map out similar regions, this is expected.  $\Sigma_{HI}$ within the optical disk also has a similar distribution, with one exception at low surface densities that we will point out later.  The similarities among the $\Sigma_{HI}$ ranges demonstrates that the HI is distributed close to uniformly within the galaxies.  
It is significant that all of the HI surface densities are below the critical surface density defined in \citet{Bigiel2008}, above which most of the cold gas is in molecular form.  We elaborate on this point below.

We compare the three calculations of SFR surface densities in the right panel of Fig. \ref{fig:sigmaHI}.  $\Sigma_{SFR}$ within the stellar disk is higher than the other measures of SFR surface density because most of the star formation is contained within the stellar disk.  $\Sigma_{SFR}$ within the HI disk is lower since a similar amount of star formation is being averaged over a much larger surface area.

\begin{figure*}[t]
\epsscale{1.0}
\plotone{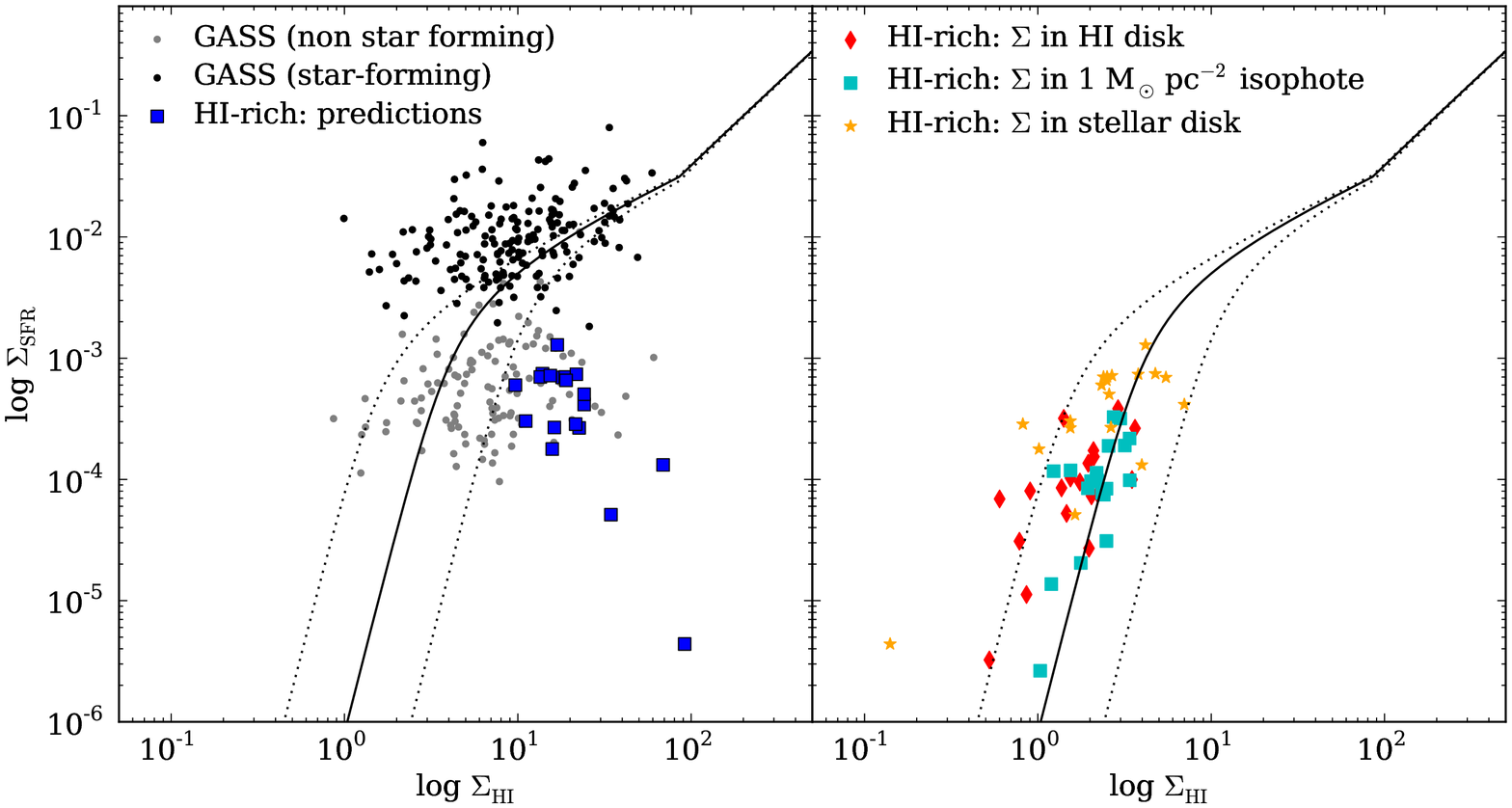}
\caption{Left panel: Predicted $\Sigma_{SFR}$ vs. $\Sigma_{HI}$ for star-forming and non star-forming galaxies in the GASS representative sample as well as the VLA sample.  We define star-forming galaxies as those with sSFRs above or within 0.5 dex of the SF sequence \citep{Schiminovich2007}.  Lines show theoretical curve for $\Sigma_{gas}$ = $\Sigma_{HI}$ + $\Sigma_{H_{2}}$ from \citet{Krumholz2009b} assuming Z = 3.16, 1.0, 0.316 (left to right).  Typical gas-phase metallicities should be solar \citep{Moran2012}.  Right panel: actual measurements of $\Sigma_{SFR}$ and $\Sigma_{HI}$ for galaxies in the VLA sample.  $\Sigma_{SFR}$ and $\Sigma_{HI}$ here are calculated within the same regions since we know the distribution of the HI.}
\label{fig:SFlaw} 
\end{figure*}

A direct comparison between SFR and HI surface densities is in Fig. \ref{fig:SFlaw}.  Since GASS galaxies have single-dish HI measurements from Arecibo, which does not provide spatial information, we cannot compare the HI surface densities for the VLA galaxies directly to the HI surface densities for more typical galaxies.  Instead, we first examine how well GASS and VLA galaxies follow theoretical predictions for the relationship between SFR and HI surface densities by assuming that their HI extents are typical, e.g. that half of the HI lies within the optical disk. 
We display these predicted surface densities for star-forming and non-star-forming galaxies from GASS and for the VLA galaxies separately in the left panel of Fig. \ref{fig:SFlaw}.  We compare them to curves for $\Sigma_{gas}$ = $\Sigma_{HI}$ + $\Sigma_{H_{2}}$ from \citet{Krumholz2009b}, which match the observations in \citet{Bigiel2008}.  The GASS galaxies lie close to the theoretical curve for galaxies with solar metallicities.  The gas-phase metallicities for these galaxies should be close to solar, although \citet{Moran2012} showed that HI-rich galaxies tend to have lower metallicities beyond R$_{90}$.  
The theoretical curves include HI and H$_2$ and we only have HI measurements.  Any measurable H$_2$ would shift the galaxies to the right in the plot.  We highlight the surface density predictions for galaxies in the VLA sample, which overlap somewhat with the non-star forming galaxies but are further from the theoretical curves than the average galaxy.  At this location, they lie below the star formation law with low SFR surface densities compared to their high HI surface densities.  If this were their true location in this plane, then we would need to appeal to factors other than low HI surface densities that can prevent gas from forming stars.

In the right panel of Fig. \ref{fig:SFlaw} we show the true locations of the HI-rich galaxies based on the VLA observations.  It is significant that the HI-rich galaxies in the VLA sample lie remarkably close to the theoretical curve for solar metallicities and in the region of the plot in which the SFR surface density drops steeply as HI surface density also decreases.  The location of the galaxies in the $\Sigma_{SFR}$ - $\Sigma_{HI}$ plane suggests that the relationship between cold gas and SFR surface densities in these galaxies conforms to theoretical expectations and that the discrepancy between their low sSFRs and high HI masses is simply due to the distribution of HI, which yields low gas surface densities.  The outlier in this plot and the only galaxy with log $\Sigma_{SFR}$ $<$ -5 is GASS 13340, which has the most extended HI disk and very little star formation.

\begin{figure*}[t]
\epsscale{1.0}
\plotone{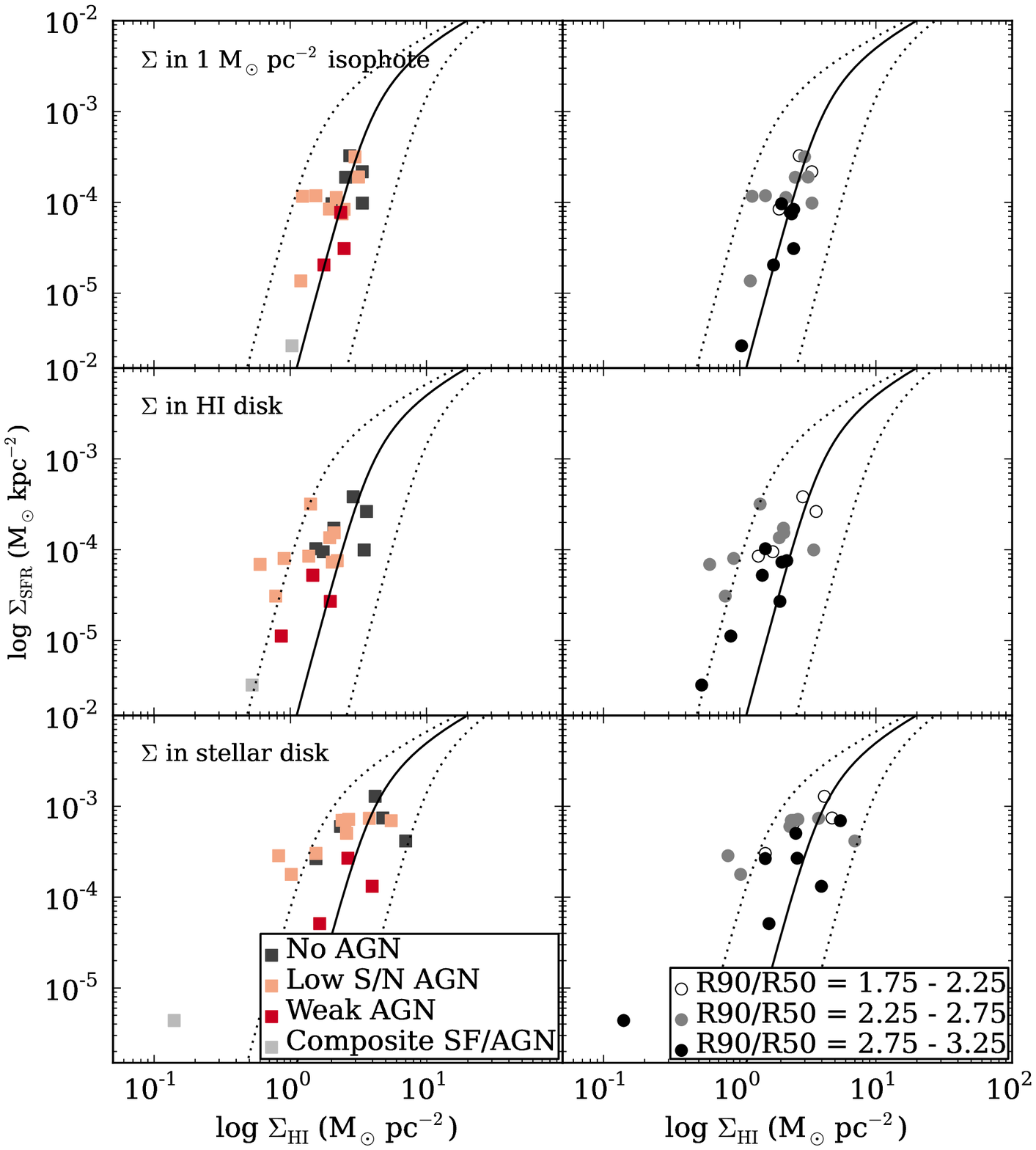}
\caption{The same data as in Fig. \ref{fig:SFlaw} with AGN classifications and concentration index (R$_{90}$/R$_{50}$) indicated.}
\label{fig:SFlaw2} 
\end{figure*}

Although it appears that low HI surface densities could be a major reason for the low total sSFRs in the VLA galaxies, other mechanisms likely contribute.  As seen in \citet{Bigiel2008} and \citet{Leroy2008}, at low gas surface densities a narrow range of gas surface densities can yield a wide range of SFR surface densities, which we also see in our sample.  To explain this phenomenon, there must exist factors other than the gas surface density that regulate the relationship between cold gas and star formation.  We consider factors related to the internal structure of the galaxy: concentration index and AGN classification.

In Fig. \ref{fig:SFlaw2} we show the $\Sigma_{SFR}$-$\Sigma_{HI}$ plane again, indicating the concentration index and AGN classification for each galaxy.  Some interesting trends are apparent, which point to the structure of the galaxy possibly playing a role in the suppression of star formation.  In the left panel we indicate whether each galaxy has no AGN, a low signal-to-noise AGN, a weak AGN (with OIII luminosity $<$ 10$^7$ L$_{\odot}$), or a spectrum indicating the presence of an AGN and star formation.  Except for GASS 13340, evidence for the presence of an AGN becomes more common towards lower SFR surface densities.  There exists a similar trend with respect to concentration, with more bulge-dominated galaxies exhibiting lower SFR surface densities.  We discuss the implications of these trends below.

\section{THE SUPPRESSION OF STAR FORMATION IN MASSIVE HI-RICH GALAXIES}
 
A number of factors regulate the relationship between cold gas and star formation, and although the HI maps move us closer to understanding the relationship between these quantities in galaxies in the VLA sample, it is still difficult to draw firm conclusions.  In the previous section we showed that low HI surface densities and the presence of a bulge might prevent HI-rich galaxies from forming stars at high rates.  These situations are commonly invoked to explain low SFRs in quenched galaxies, but we emphasize that our galaxies are not ``quenched" in the traditional sense: they have a surplus of cold gas and they are still forming stars, albeit at a low rate.  In this section we delve into the observational and theoretical evidence for these star formation suppression mechanisms to understand not just if these mechanisms are acting in these galaxies but how we might be able to use additional data to draw firmer conclusions.
 
We discuss in detail how AGN feedback and the presence of a bulge can suppress star formation and we describe why galaxies in the VLA sample might harbor long-lived low-surface-density HI disks.  We also explain why recent gas accretion probably cannot account for the extreme HI masses in these galaxies.  Finally, we discuss why galaxies with an unexpected combination of high HI masses and low sSFRs become common only at high stellar masses.  We do not consider here environmental quenching mechanisms, such as ram-pressure stripping \citep[e.g.][]{Tonnesen2007}, which remove substantial amounts of gas from galaxies, though it would be interesting to see if galaxies in the VLA sample are preferentially in extremely isolated environments, such as voids, where gas-rich galaxies are found with a higher frequency \citep{Kreckel2012}.  We do know that the HI-rich galaxies are not cluster galaxies, but a thorough examination of their environments and a possible link between their environments and their gas and star-forming properties is beyond the scope of this work.

\subsection{AGN Feedback}
AGN can inhibit star formation in at least three distinct ways, which can occur simultaneously but in different proportions within a given galaxy.  AGN can remove potentially star-forming gas from galaxies via outflows; the energy released by AGN can contribute to the existence of a hot halo, preventing future gas accretion; and AGN can inject energy into the gas reservoir such that it is unable to collapse and form stars but stays within the galaxy.  Most observational studies of star formation suppression via AGN focus on the detection of outflows as evidence that AGN can remove enough gas to significantly affect a galaxy's SFR \citep[e.g.][]{CanoDiaz2012, Yuma2013, Forster2013}, 
but this type of suppression via AGN cannot explain the large cold gas contents in the HI-rich galaxies.  The buildup of a hot halo that prevents gas accretion cannot explain why the cold gas already within the galaxy is stable against star formation, so is not helpful in explaining our results.  The third scenario, in which AGN feedback disrupts the cold gas enough to prevent it from forming stars, could explain our results and is the scenario we discuss below.

Because simulations have only just begun to include a multiphase ISM \citep[e.g.][]{Dave2013}, to our knowledge none exist that specifically address how the HI that remains in a galaxy is affected by AGN feedback (though \citet{Gabor2013} note that in gas-rich z$\sim$2 galaxies AGN feedback is not strong enough to disrupt dense clouds of star-forming gas).  Observational studies of AGN hosts with measured HI masses show that HI can remain at high levels in galaxies with AGN: \citet{Ho2008} and \citet{Fabello2011b} show that AGN hosts contain just as much HI as do their counterparts without AGN.  If massive galaxies can simultaneously contain AGN and large HI reservoirs, how, if at all, is the AGN affecting the HI reservoir?  Direct observational evidence of this is lacking, but \citet{Nesvadba2010} show how AGN feedback prevents the \emph{molecular} gas from forming stars in the H$_2$-luminous radio galaxy 3C 326 N.  Like the HI-rich galaxies in this paper, 3C 326 N contains a substantial amount of cold gas (M$_{H2}$ = 10$^9$ M$_{\odot}$) but a surprisingly low SFR.  
They found that mechanical energy from the AGN that was injected into the ISM heated the molecular gas enough to make it stable against star formation but still detectable as H$_2$.

It is possible that a similar mechanism that heats the HI is at work in some of the massive HI-rich galaxies discussed here.  If this is so, HI-rich galaxies with AGN should have a lower SFR surface density than comparison galaxies with the same HI surface density.  In Fig. \ref{fig:SFlaw2} we show that this is indeed the case: galaxies with lower SFR surface densities tend to have evidence of an AGN.  However, the only true AGN in the VLA sample are weak, with OIII luminosities less than 10$^{7}$ L$_{\odot}$, so it is unclear if they are powerful enough to affect the large quantities of gas in galaxies in the VLA sample.  

\subsection{Morphological Quenching}

One way in which star formation can be suppressed without removing or heating the cold gas is via morphological quenching, in which a galaxy's bulge stabilizes its gas disk against star formation.  \citet{Martig2009} explain that gas disks embedded in bulge-dominated galaxies may exhibit SFRs a factor of 10 lower than similarly massive gas disks in spiral galaxies for two reasons: bulge-dominated galaxies lack a stellar disk that contributes to the self-gravity of a gas disk and naturally have higher epicyclic frequencies, which can increase the Toomre Q parameter above the critical value for star formation.  This scenario is also seen in simulations at redshift z$\sim$2 \citep{Agertz2009, Ceverino2010} and is supported by observations at z$\sim$2 in which star-forming galaxies have higher values of Q at their centers where a bulge dominates \citep{Genzel2013}.  In the local universe, lower HI and H$_2$ SFEs in bulge-dominated galaxies compared to late-type galaxies provide some evidence for morphological quenching \citep{Saintonge2012}.

\citet{Martig2013} find that the effectiveness of morphological quenching depends on the precise gas content of the galaxy.  Early-type galaxies with low cold gas fractions (1.3\% in their simulation) contain gas disks that are stable against star formation and do not fragment.  Early-type galaxies with higher gas fractions (4.5\% in their simulation) have gas disks that do fragment but form stars less efficiently than similar gas disks embedded in a spiral galaxy because a smaller fraction of their cold gas is in a very dense ($>$ 10$^4$ cm$^{-2}$) phase.  The reduced fragmentation due to morphological quenching effectively lowers the fraction of cold gas in dense phases, which in turn lowers the global efficiency of star formation.  Taking into account the range of possible gas fractions, \citet{Martig2013} report that gas disks embedded in bulge-dominated galaxies form stars 2-5 times less efficiently than do gas disks with a similar gas surface density embedded in spirals.  

While a direct comparison between the \citet{Martig2013} simulations and our HI-rich galaxies is impossible because the HI-rich galaxies in this paper have much higher gas fractions, ranging from 15\% to over 60\%, their results suggest that bulges could play a role in lowering the SFE in some of our HI-rich galaxies.  \citet{Martig2013} emphasize that morphological quenching is not strong enough to drive the evolution of all galaxies onto the red sequence, but could be a contributing factor alongside other quenching mechanisms. 

\citet{Martig2013} show in the radial profiles of their simulated galaxies that the gas surface density increases to greater than 10$^2$ M$_{\odot}$ pc$^{-2}$ at the centers of the galaxies.  The gas in this density regime is likely molecular, which would create a central hole in a map of the HI flux.  Some galaxies in the VLA sample do exhibit an HI depression in the center.  If galaxies with little HI in their centers do have central H$_2$ and no sign of central star formation, this could be evidence of morphological quenching.  If the center lacks cold gas altogether, it is unclear if morphological quenching is at work.  H$_2$ observations of galaxies in the VLA sample could bring clarity to this scenario.

Since all of the HI-rich galaxies described here have high gas fractions, which \citet{Martig2013} find lower the effect of morphological quenching, and most have low gas surface densities, which according to \citet{Martig2013} heighten the effect of morphological quenching, we cannot draw any definitive conclusions about the effect of morphological quenching in our sample except to say that it could be playing a role in the more highly-concentrated galaxies based on the recent simulations and observations demonstrating its effects.  In Figure \ref{fig:SFlaw2} we identify the HI-rich galaxies based on their concentration index; there is some evidence that the more bulge-dominated galaxies have lower SFR surface densities at a given HI surface density, though it is unlikely that morphological quenching can affect the very extended gas disks in the VLA sample.  Morphological quenching is unlikely to play a role in less bulge-dominated galaxies.

\subsection{Below-Threshold Cold Gas}

Inefficient star formation can result from low cold gas surface densities, which can be present in HI-rich galaxies if the HI extends over a large surface area.  As noted above, HI disks generally extend well beyond stellar disks, and this is something we see in the VLA sample as well.  Galaxies selected by HI tend to be more extended than those that are selected optically \citep{Huang2012}.

Extended HI disks will form stars inefficiently only if the disks do not contain substantial amounts of H$_2$, the immediate precursor to star formation.  Observations confirm that the outskirts of gaseous disks tend to be HI-dominated.  \citet{Bigiel2008} show that in local spirals, HI extends well beyond the radius at which SFR and H$_2$ become negligible.  
In \citet{Leroy2008} the molecular fraction, $\Sigma_{H_2}$/$\Sigma_{HI}$, decreases steadily with radius, and in \citet{Martin2001} the total (HI + H$_2$) gas surface density profile decreases more quickly than the HI surface density profile.  \citet{Schaye2004} shows that extended gas disks tend to be HI-dominated because the transition from the neutral to the molecular phase proceeds efficiently only above a critical gas surface density of 3-10 M$_{\odot}$ pc$^{-2}$.  A critical surface density of 9 M$_{\odot}$ pc$^{-2}$ is confirmed observationally by \citet{Bigiel2008}, above which cold gas is mostly molecular.

Our observed galaxies have HI surface densities that fall below the critical gas surface density, indicating that the gas disks in the VLA sample are likely HI-dominated.  Where HI dominates over H$_2$, star formation necessarily proceeds at a slower rate.  Observations show that in the outskirts of galaxies where HI dominates, and in dwarfs that are HI-dominated throughout, the local SFE decreases with radius \citep{Bigiel2008, Leroy2008}.  Integrated and local measurements show that there is a break in the star formation law such that at gas surface densities below the critical density, galaxies exhibit SFR surface densities a factor of 5 below the extrapolation of the star formation law at higher gas surface densities \citep{Wyder2009, Bigiel2008}.  \citet{Wyder2009} show that this downturn, which is equivalent to a lowering of the SFE at low gas surface densities, agrees with work by \citet{Krumholz2005} and \citet{Blitz2006} predicting lower H$_2$ fractions in low surface density gas.  The galaxies in the VLA sample lie in the part of the star formation law in which star formation is least efficient.  Thus, the low sSFRs in the galaxies in the VLA sample are reasonable given their HI surface densities.

For the phenomenon of extended, low surface density HI disks to be common, the extended HI disks must be long-lived.  Simulations have shown that extended low surface density gaseous disks may result when galaxies accrete from the IGM cold, dense filaments, which have higher angular momenta than the dark matter halo and prevents efficient gas transport to the disk \citep{Kimm2011, Stewart2011}.  A more likely scenario for the massive galaxies in the VLA sample, which might not accrete gas directly from cold filaments in the IGM, is described in \citet{Lu2014}.  They show that extended gas disks with HI surface densities below the critical density for star formation can result when the gas disks are built up by gas cooling from large radii within the hot halo.  

Populations of galaxies in the local universe that are defined by inefficient outer star formation include GLSBs and XUV-disks.  Galaxies in the VLA sample might be analogs of GLSBs: in Fig. 1 we show that GLSBs and the VLA sample lie in the same region of the HI gas fraction - stellar mass plane.  The prevalence of XUV-disks \citep{Thilker2007, Lemonias2011} confirms that inefficient star formation in extended disks is common.  Galaxies in the VLA sample are probably not XUV-disks (except for 45664) because their UV images do not show evidence of recent star formation within the extended HI, but galaxies in the VLA sample and XUV-disks might represent different stages of the life of an extended HI disk.

\subsection{Recent Gas Accretion}

If galaxies in our sample recently accreted cold gas, it is possible that the gas simply hasn't been in the galaxy long enough to collapse and form stars.  \citet{Keres2005} show that it generally takes 0.5 Gyr for the cosmic SFR to react to new gas accretion.  To determine whether this scenario could account for the high HI masses and low SFRs of galaxies in our sample, we must consider, first, whether we expect galaxies at this mass scale to be accreting gas and at what rate, and, second, whether we see any evidence of recent accretion in the optical images and HI maps.  

Simulations have shown that the ways in which galaxies accrete gas depend on their mass, environment, and redshift.  At z$\sim$0, gas that is accreted onto galaxies above a critical halo mass is shock heated to the virial temperature and likely stays warm \citep{Birnboim2003, Dekel2006, Keres2005}.  Gas that is accreted onto galaxies below the critical halo mass tends to enter a galaxy in the form of cold streams from the IGM.  Depending on the precise way in which one relates halo mass to baryonic mass, this critical mass could lie just below the stellar mass at which many HI-rich galaxies with low SFRs are found.  Thus, galaxies like those in our sample are probably accreting gas that is shock-heated and it is unclear whether that will ultimately cool into HI.  

For recent accretion to account for the large amount of excess HI in our HI-rich galaxies, we would expect to see large tidal features indicative of major accretion events in the HI maps or optical imaging.  We do not see any obvious signatures of accretion events in the VLA sample and the HI maps suggest that galaxies in the VLA sample have regularly rotating, settled disks.  GASS 45664 is the possible exception, with an HI distribution whose peak is offset from the stellar disk.  The regions with the strongest HI emission exhibit prominent spiral arms, but other regions with significant HI do not have signs of star formation.  This combination of UV-bright features embedded in an extended HI disk that does not show signs of star formation elsewhere suggests that we are seeing this galaxy in the process of re-building its stellar disk after a major accretion event.  

In this initial study of HI morphologies, we cannot assess the distribution and kinematics of the HI at the level done by \citet{Sancisi2008} because our observations do not have the required depth.  That type of analysis could reveal lesser accretion events based on the presence of gas at anomalous velocities, but such lesser accretion events are unlikely to yield the high HI masses in the VLA sample.  
It appears that much of the cold gas in galaxies in the VLA sample has been retained as the galaxies evolved since there is no evidence that major episodes of accretion brought in large amounts of cold gas.

\subsection{The Emergence of Low-Star-Forming HI-Rich Galaxies at High Stellar Masses and the Role of H$_2$}
\label{sec:H2}

The results of our analysis of HI-rich galaxies in Section 2 are twofold:  1) that a population of HI-rich galaxies with surprisingly low sSFRs exists, and 2) that this population primarily exists above a stellar mass of log M$_*$ = 10.5.  HI imaging allows us to understand the suppression of star formation in these galaxies.  A separate question is why galaxies with an excess of HI compared to their SFRs become common above a threshold stellar mass. 

Some evidence exists to show that as stellar mass increases, not only do sSFRs decrease, but so does the efficiency with which galaxies convert their cold gas into stars.  \citet{Young2013} show that cold gas (HI or H$_2$) content up to 10$^9$ M$_{\odot}$ does not measurably affect the UV-optical colors of massive (log M$_*$ $>$ 10.7) galaxies.  They also find that less massive early-type galaxies with cold gas are blue while more massive early-type galaxies with measurable cold gas are red.  Even when massive galaxies have significant amounts of H$_2$, which is one step closer to forming stars than HI, they are unlikely to form stars at a high rate.  Although the HI-based SFE is constant with respect to stellar mass \citep{Schiminovich2010}, the H$_2$-based SFE decreases as stellar mass increases \citep{Saintonge2011b}.  \citet{Young2013} show that massive galaxies have older single stellar population ages than less massive galaxies with the same H$_2$ fraction.  Together, these findings present a picture in which the most massive galaxies do not form stars even if they have the cold gas necessary to do so.  

A significant HI mass compared to SFR could be a sign that a galaxy is lacking H$_2$ and is suffering from inefficient conversion from HI to H$_2$.  \citet{Huang2012} suggest this is why their HI-selected sample has lower SFEs than an optically selected sample.  An abundance of HI compared to H$_2$ might be more common in more massive bulge-dominated galaxies.  \citet{Saintonge2011a} find that the detection rate of HI and H$_2$ drops significantly among more bulge-dominated galaxies and that above these detection thresholds, galaxies have mostly HI if they have any cold gas at all.  Similarly, red galaxies are more likely to be detected in HI than H$_2$  \citep{Young2013}.  

\citet{Lee2014} measure the H$_2$ in 28 HI-rich galaxies with stellar masses and HI masses similar to those for our sample.  15 out of 20 normal galaxies and 4 out of 8 LSB galaxies were detected.  Their high H$_2$ contents place them at the high end of the H$_2$ gas fraction distribution shown in \citet{Saintonge2011a}, but their cold gas reservoirs are still dominated by HI over H$_2$ by a factor of 2 to 3.  Although our sample was selected differently from theirs to include only HI-rich galaxies with low sSFRs, there is still no evidence that HI-rich galaxies in general are molecular-dominated.  

Although we do not have H$_2$ measurements for the VLA sample, we assume based on the results of \citet{Saintonge2011a, Young2013, Lee2014} 
and on the low sSFRs in these galaxies that they do not contain significant quantities of H$_2$.  What, then, is hindering the conversion from HI to H$_2$ in massive galaxies?  \citet{Fu2010} predict that a high spin parameter $\lambda$ (usually associated with more massive galaxies) is related to less efficient HI-to-H$_2$ conversion.  \citet{Huang2012} show that at a given stellar mass, galaxies with higher HI fractions are in dark matter halos with higher spin parameters, which probably means that they are more extended since their halos have higher angular momenta.  Although this phenomenon is no longer obvious at stellar masses above M$_{*}$ $\sim$ 10.5, \citet{Huang2012} generally find that HI-selected galaxies reside in halos with high spin parameters and are more extended.  If massive HI-rich galaxies tend to have more extended HI disks than is typical, then the conversion from HI to H$_2$ must be weak because much of the gas lies beyond the radius at which the conversion proceeds efficiently \citep{Schaye2004}.

\section{Summary and Conclusions}

We used two recent large HI surveys, GASS and ALFALFA, to define and select a sample of massive galaxies (log M$_*$/M$_{\odot}$ $>$ 10) that are extremely HI-rich for their stellar mass.  The HI-rich galaxies have HI fractions in the top 5\% for their stellar mass and have HI masses greater than 10$^{10}$ M$_{\odot}$.  Within the HI-rich sample, we examined the relationship between HI and star formation as a function of stellar mass.  We found that even though the HI fractions of the HI-rich galaxies decrease with stellar mass, their sSFRs decrease at a stronger rate.  This trend of decreasing sSFRs revealed a sample of HI-rich galaxies with surprisingly low sSFRs at the high end of the stellar mass range (log M$_*$/M$_{\odot}$ $>$ 10.5).  

To understand the physical conditions producing this unexpected combination of very high HI masses and lower sSFRs, we obtained HI maps at the VLA of 20 of these galaxies.  The HI maps yield the distribution of the HI along with the extent and surface density of the HI within the HI disk.  We found that the HI surface densities are low enough that the galaxies fall in the region of the $\Sigma_{SFR}$-$\Sigma_{HI}$ plane in which inefficient star formation is common.  In this regime, a narrow range of HI surface densities can yield a wide range of SFR surface densities.  

Because the HI surface densities for the galaxies in the VLA sample are low, their low sSFRs are not unexpected.  But other conditions, including the internal structure of the galaxies, could also contribute to their low sSFRs.  We found that galaxies with the lowest SFR surface densities are more likely to be bulge-dominated and exhibit stronger evidence of AGN.  However, because bulge-dominated galaxies are generally more likely to host AGN, it is unclear whether the AGN or the structure of the galaxy (in the form of morphological quenching) is contributing to the suppression of star formation.  Future observations of the molecular gas in these galaxies could provide some insight.  A more detailed morphological analysis of the HI akin to that in \citet{Holwerda2011} could also provide some clues, but the distances to these galaxies coupled with the short integration times of the observations described here make that type of work less robust.

It is not obvious that the same star formation suppression mechanisms must operate in other types of galaxies with less HI, but it is worth considering how these results might apply to other populations of galaxies with low sSFRs.  Since all of the galaxies in the VLA sample have low HI surface densities, there is no strong evidence that high concentration index or AGN are driving the low sSFRs, but they seem to be contributing.  As we have shown, ascertaining the driving factor behind suppressed star formation is not straightforward even when maps of the cold gas are available.  Nevertheless, it could be worth examining in further detail how the HI distributions and HI surface densities vary with stellar mass within populations of galaxies with less cold gas to see if the same conditions apply in less extreme galaxies.  Of course some massive galaxies have low sSFRs because they lack the cold gas necessary for star formation to proceed, but we have shown that star formation drops with stellar mass even among populations of galaxies that have extremely high quantities of HI, so other conditions must be at work in the broader population.

Although galaxies in the VLA sample appeared to challenge the star formation law with their unexpected combination of high HI masses and low sSFRs, the sample actually conforms to the global star formation law.  We have shown that the star formation law is an important tool for understanding the relationship between gas and star formation in galaxies that appear to be outliers.  Placing other atypical galaxies, such as XUV-disks, GLSBs, or galaxies transitioning between the red and blue sequences, on the $\Sigma_{SFR}$-$\Sigma_{HI}$ plane could improve our understanding of these important populations.

\appendix

\section{Notes on Individual Galaxies}

\textbf{13340}  Red galaxy with faint extended features.  Double-horned profile in ALFALFA.  Very faint signal in the UV.

\textbf{15607}  Inclined, possible S0.  Blue, edge-on galaxy less than 1 arcminute away with no redshift.  Double-horned profile in ALFALFA.  

\textbf{17705}  Face-on spiral with possible ring and faint blue outer structure.  This galaxy is at z=0.035; reddish spiral 2 arcminutes away at z=0.036.  Very narrow, highly peaked double-horn profile in ALFALFA. 

\textbf{19918}  Broad ALFALFA spectrum with one narrow peak on one side.

\textbf{23187}  Tightly wound spiral.  Lopsided double-horned profile in ALFALFA.  Strong UV flux extends to edge of faint optical disk.

\textbf{25285}  Tightly wound spiral with blue arms.  Double-horned profile in ALFALFA.  Two lobes of HI on each side of galaxy.  Half of VLA data were corrupted.

\textbf{26806}  Tightly wound spiral with at least one faint, blue, extended arm.  Very strong double-horn structure in ALFALFA spectrum.  Strong UV flux extends beyond main stellar body.

\textbf{29304}  Reddish main body with faint blue outer arms.  Double-horned spectrum in ALFALFA.

\textbf{40245}  Reddish featureless galaxy with faint extended structure and one prominent blue tail/arm.  Wide double-horned spectrum in ALFALFA.  UV flux coincides only with main stellar body.

\textbf{41063}  Reddish tightly wound spiral. UV flux coincides only with main stellar body.  

\textbf{42262}  Inclined spiral.  Wide ALFALFA spectrum with peak only at low-velocity end.

\textbf{45267}  Reddish center with blue flocculent spiral arms in outskirts of galaxy.  Double-horned profile in ALFALFA.

\textbf{45357}  Inclined spiral.  Broad and faint ALFALFA spectrum.  Strong UV flux through main stellar body.  Two HI lobess on each side of galaxy, maybe indicative of an HI ring

\textbf{45404}  Ring-like galaxy with very extended diffuse light.  Blue arm of star formation 20 arcsec away from galaxy and unattached (in optical).  Double-horned profile in ALFALFA.  Best spectral window had bad data; very close to edge of spectral window used here.  

\textbf{45664}  Featureless elliptical with two very long blue arms that are strong in the UV but appear unattached to main galaxy at optical wavelengths.   More bright, patchy UV surrounds galaxy.  This galaxy is at z=0.0364; red barred spiral 1 arcmin to the north at z=0.038.  Another at 2 arcminutes at z=0.038.  Broad, uneven ALFALFA spectrum, sloping upwards at higher velocities.  Several bright patches of HI, many of which coincide with bright patches in UV.

\textbf{47677}  Featureless elliptical at optical wavelengths.  Very little UV flux.  Blue edge-on galaxy 3 arcminutes away at same redshift.  Red edge-on galaxy 2.5 arcminutes away at similar redshift.  Reddish inclined galaxy 1 arcmin away at similar redshift.  Faint blue spiral very nearby at same redshift.  Is this a group or overdensity?  Lopsided ALFALFA spectrum with more flux at low velocities.  

\textbf{47708}  Edge-on galaxy with strong dust lane.  Very little UV flux.  Broad and faint ALFALFA spectrum.  Prominent HI structure offset from but in line with galaxy.

\textbf{51390}  Early-type barred spiral.  UV flux extends beyond main stellar body.  Triple-peaked spectrum in ALFALFA.

\textbf{57949}  Spiral with tightly wound blue arms and knot of star formation at the tip of an arm.  UV flux coincides with optical flux.  Double-horned profile in ALFALFA, with much stronger peak at higher velocities.  

\acknowledgments
We thank Ximena Fern\'{a}ndez and Jacqueline van Gorkom for sharing their expertise and Ted Wyder for providing the GLSB data.  We also thank the anonymous referee for useful comments.  J.L. was partially supported by HST-GO-12603.02-A and a NASA Space Grant.  B.C. is the recipient of an Australian Research Council Future Fellowship (FT120100660).  

The National Radio Astronomy Observatory is a facility of the National Science Foundation operated under cooperative agreement by Associated Universities, Inc.

The Arecibo Observatory is operated by SRI International under a cooperative agreement with the National Science Foundation (AST-1100968), and in alliance with Ana G. M\`{e}ndez-Universidad Metropolitana, and the Universities Space Research Association.

\emph{GALEX (Galaxy Evolution Explorer)} is a NASA Small Explorer, launched
in April 2003. We gratefully acknowledge NASA's support for
construction, operation, and science analysis for the \emph{GALEX} mission,
developed in cooperation with the Centre National d'Etudes Spatiales
(CNES) of France and the Korean Ministry of Science and Technology.

This work has made extensive use of the \url[]{MPA/JHU}
SDSS value-added catalogs.

Funding for the SDSS and SDSS-II has been provided by the Alfred
P. Sloan Foundation, the Participating Institutions, the National
Science Foundation, the U.S. Department of Energy, the National
Aeronautics and Space Administration, the Japanese Monbukagakusho, the
Max Planck Society, and the Higher Education Funding Council for
England. The SDSS Web Site is http://www.sdss.org/.

The SDSS is managed by the Astrophysical Research Consortium for the
Participating Institutions. The Participating Institutions are the
American Museum of Natural History, Astrophysical Institute Potsdam,
University of Basel, University of Cambridge, Case Western Reserve
University, University of Chicago, Drexel University, Fermilab, the
Institute for Advanced Study, the Japan Participation Group, Johns
Hopkins University, the Joint Institute for Nuclear Astrophysics, the
Kavli Institute for Particle Astrophysics and Cosmology, the Korean
Scientist Group, the Chinese Academy of Sciences (LAMOST), Los Alamos
National Laboratory, the Max-Planck-Institute for Astronomy (MPIA),
the Max-Planck-Institute for Astrophysics (MPA), New Mexico State
University, Ohio State University, University of Pittsburgh,
University of Portsmouth, Princeton University, the United States
Naval Observatory, and the University of Washington.


 \end{document}